\documentclass[pdftex,numberedappendix,appendixfloats,apj,twocolappendix]{emulateapj}
\usepackage{amsmath}

\begin{document}

\shorttitle{Metals in dwarfs in tidal fields}

\shortauthors{Williamson, Martel, \& Romeo}

\title{Chemodynamic evolution of dwarf galaxies in tidal fields}

\author{David Williamson}
\affil{D\'epartement de physique, de g\'enie physique et d'optique, Universit\'e Laval, Qu\'ebec, QC, G1V 0A6, Canada}
\affil{Centre de Recherche en Astrophysique du Qu\'ebec, QC, Canada}
\email{david-john.williamson.1@ulaval.ca}

\author{Hugo Martel}
\affil{D\'epartement de physique, de g\'enie physique et d'optique, Universit\'e Laval, Qu\'ebec, QC, G1V 0A6, Canada}
\affil{Centre de Recherche en Astrophysique du Qu\'ebec, QC, Canada}

\and

\author{Alessandro B. Romeo}
\affil{Department of Earth and Space Sciences, Chalmers University of Technology, SE-41296 Gothenburg, Sweden}

\makeatletter{}\begin{abstract}

The mass-metallicity relation shows that the galaxies with the lowest mass have the lowest metallicities. As most dwarf galaxies are in group environments, interaction effects such as tides could contribute to this trend. We perform a series of smoothed particle hydrodynamics (SPH) simulations of dwarf galaxies in external tidal fields to examine the effects of tides on their metallicities and metallicity gradients. In our simulated galaxies, gravitational instabilities drive gas inwards and produce centralized star formation and a significant metallicity gradient. Strong tides can contribute to these instabilities, but their primary effect is to strip the outer low-metallicity gas, producing a truncated gas disk with a large metallicity. This suggests that the role of tides on the mass-metallicity relation is to move dwarf galaxies to higher metallicities.

\end{abstract}

\keywords{
galaxies: interactions --- galaxies: abundances --- galaxies: dwarf --- galaxies: evolution
}

\maketitle

\makeatletter{}
\section{Introduction}

It is well established that there is a tight correlation between the mass and metallicity of galaxies \citep[e.g.][]{2004ApJ...613..898T,2013MNRAS.430.2680M}. This correlation provides an explanation for the weaker luminosity-metallicity relation \citep{1987ApJ...317...82G,1989ApJ...347..875S,1991ApJ...379..157B,1994ApJ...420...87Z}. These correlations demonstrate that more massive/luminous galaxies have higher metallicities than less massive/luminous galaxies. Investigations into the chemodynamics of galaxies can yield insight on this trend.

Clearly, metallicity is either preferentially enhanced in more massive galaxies, or depleted in less massive galaxies. Metal enrichment is a result of star formation and evolution, and could explain the mass-metallicity relation if more massive galaxies had a disproportionately large amount of star formation for their gas mass. On the other hand, observational studies show globally high specific star formation rates and short depletion times in dwarf galaxies, and a trend for specific star formation rates to decrease with mass \citep[e.g.][]{2011ApJ...739L..40R,2012ApJ...754L..29W,2013AJ....146...19L,2014MNRAS.443.1329H,2016A&A...590A..27G,2016ApJ...820..109F}. Thus, star formation appears to be more efficient in dwarf galaxies, even if local sub-kpc sized regions of dwarfs can have extremely low star formation efficiencies \citep{2008AJ....136.2846B}, similar to those of the outer disks of larger galaxies \citep{2008AJ....136.2846B,2010AJ....140.1194B}. Accretion of pristine gas could reduce the metallicity of a galaxy, but can also act to ``refuel'' star formation and produce additional metals \citep{2012MNRAS.423.2991V,2014MNRAS.438..262P,2014MNRAS.442.1830V,2016MNRAS.457.2605C}, reducing the strength of this effect. Instead, metal-rich outflows that increase in efficiency as galaxy mass decreases are considered the most likely origin of the mass-metallicity relation \citep[e.g.][]{1979MNRAS.186..503T,1987A&A...173...23A,2004ApJ...613..898T,2008MNRAS.387..577O}. This effect must be at its strongest in the smallest galaxies, making the chemodynamics of low-mass galaxies the critical factor in explaining the mass-metallicity relation.

Hence there is motivation to investigate the chemodynamics of low-mass galaxies in detail. However, there is disagreement as to why outflows may be more effective in these galaxies. It has been argued \citep{1986ApJ...303...39D,1987A&A...173...23A} that this results from the relative depth of the potential wells. This is possible if the outflow energy couples well to the ambient halo gas \citep{2003MNRAS.344.1131D}, but \citet{2008MNRAS.385.2181F} have shown this is not the case. Hydrodynamic effects -- specifically, confinement from infalling gas \citep{2005ApJ...634L..37F,2008MNRAS.387..577O} may also prevent enriched gas from being lost into the intergalactic medium (IGM). Many dwarf galaxies are in groups, where encounters should be common. Ram-pressure and tidal stripping \citep{2006MNRAS.369.1021M,2008ApJ...672L.103K,2013MNRAS.436.1191T} may also contribute to removing enriched outflows from dwarf galaxies.

These processes will also affect the metallicity gradients of galaxies. Metallicity gradients have been observed in a number of galaxies \citep[e.g.][]{1994ApJ...420...87Z,2010ApJS..190..233M,2012ApJ...745...66M}, and their slopes are a result of a combination of the star formation gradient, the strength and scale of galactic outflows and inflows, and the time and length scales of gas mixing and stellar migration. A positive metallicity gradient could be steepened or even inverted by the accretion of pristine gas \citep{2010Natur.467..811C,2012ApJ...745...66M,2016MNRAS.457.2605C}, steepened by centralized star formation \citep{2012A&A...540A..56P}, and flattened by a feedback-driven galactic fountain spreading out metals \citep{1997ApJ...478L..21M,2003MNRAS.340..908K,2013A&A...554A..47G,2013MNRAS.434.1531F}, or by gas mixing \citep{2012ApJ...750..122B,2015MNRAS.449.2588P}, and by stellar migration \citep{2010ApJ...722..112M,2013A&A...553A.102D,2014MNRAS.439..623G}.

While real galaxies are likely to be affected by a combination of many of these effects, it is useful to examine individual processes in isolation, to reduce the degeneracy and model-dependence that comes with a more comprehensive treatment, and to investigate weaker effects that may be ``washed out'' by more dominant processes. Here, we solely consider the effects of a tidal field on the chemodynamic evolution of a dwarf galaxy. We do not seek to explain the entire mass-metallicity relation, but focus on one aspect of the chemodynamical evolution of the most critical galaxies.

Tides could induce several effects. Tidal stripping can remove gas from the galaxy, especially from the outer regions \citep[e.g.][]{1994A&ARv...6...67F} (which may be metal-poor if the metallicity gradient is negative), or from extended outflows. Tidal stirring from a varying field could induce gravitational instabilities that drive gas inwards and provide fuel for star formation and metal production \citep[e.g.][]{1991ApJ...381...14L}. Tides could potentially stabilize against such instabilities and lower the star formation rate, if they increase the scale length and velocity dispersion of the gas, increasing the Toomre Q parameter \citep{1964ApJ...139.1217T}. In principle, tides could also spread out the disk material and thus flatten the metallicity gradient of a galaxy. In this work, we examine and quantify the role of these processes on the abundance and distribution of metals in dwarf galaxies.

The rest of the paper is organized as follows: In section~\ref{methods_sec} we describe our numerical algorithm and our simulations. In section~\ref{genev} we describe the general evolution of our models. We examine the results of the simulations in more detail in sections~\ref{starformsection}--\ref{massmetsection}, describing star formation and gravitational instabilities in section~\ref{starformsection}, tidal stripping in section~\ref{stripsection}, metallicity gradients in section~\ref{flatsection}, and the dwarf galaxies in the context of the mass-metallicity relation in section~\ref{massmetsection}. We discuss the robustness of our results in section~\ref{discsection}, and summarize our conclusions in section~\ref{concsection}.

\makeatletter{}\section{Method and Models}\label{methods_sec}

We perform simulations with \textsc{GCD+} \citep{2003MNRAS.340..908K,2012MNRAS.420.3195B,2013MNRAS.428.1968K,2014MNRAS.438.1208K}, an MPI N-Body+SPH code that includes self-consistent stellar feedback, radiative cooling, and explicit metal production and diffusion, separately tracking the abundances of $^1$H, $^4$He, $^{12}$C, $^{14}$N, $^{16}$O, $^{20}$Ne, $^{24}$Mg, $^{28}$Si, and $^{56}$Fe. Details of the latest version of this code can be found in \citet{2014MNRAS.438.1208K}. We modified the code as detailed in \citet[][hereafter Paper \textsc{I}]{2016ApJ...822...91W}, and summarize the changes here.

Firstly, feedback particles directly deposit their metals to nearby particles with a weight determined by the smoothing kernel. This is justified as these are the same particles that are receiving kinetic energy from the feedback particle (through pressure). This should contribute to driving metal-rich winds.

Secondly, we damp star formation for the first $0.2$ Gyr of evolution. The star formation law in \textsc{GCD+} is 
\begin{equation}
\dot{\rho}_\mathrm{SF}^{\phantom X} = \frac{\epsilon_* \rho_g}{\tau_g},
\end{equation} where $\dot{\rho}_\mathrm{SF}^{\phantom X}$ is the star formation density, $\rho_g$ is the gas density, $\tau_g$ is the local dynamical time, and $\epsilon_*$ is the star formation efficiency. To damp star formation, we replace $\tau_g$ with $\tau_{g,\mathrm{damped}}$ for the first $0.2$ Gyr of the simulation, defining $\tau_{g,\mathrm{damped}}$ by
\begin{equation}
\tau_{g,\mathrm{damped}}=(0.2 \mathrm{Gyr}/t) \tau_g,
\end{equation}where $t$ is the time since the start of the simulation. A galaxy that evolves from the initial conditions of an axisymmetric disk in hydrodynamic equilibrium can cool and collapse uniformly, producing a rapid burst of star formation. Damping the initial star formation rate suppresses this burst.

Most critically, we have modified the method for calculating the diffusion coefficient. Metal diffusion is performed using the ``Shear'' model of Paper \textsc{I}, which calculates the diffusion coefficient using the method of \citet{2010MNRAS.407.1581S}, but otherwise solves the diffusion equation according to the method of \citet{2009MNRAS.392.1381G}. Here, the diffusion coefficient, $D_i$, of particle $i$ is given by $D_i = c_D |S| h_i^2$, where $S$ is the trace-free shear tensor, $h_i$ is the smoothing length, and $c_D$ is a scaling constant, typically in the range of $0.05-0.1$. Here, we set $c_D=0.1$.

This diffusion model is a variation on the standard eddy-viscosity method, where the diffusion coefficient is $D\approx VL$, for some velocity and length scales $V$ and $L$, which is often used in SPH simulations in similar contexts \citep[][and Paper \textsc{I}]{2009MNRAS.392.1381G,2010MNRAS.407.1581S,2013MNRAS.434.3142A,2014MNRAS.438.1208K}. Our model follows \citet{2009MNRAS.392.1381G} by selecting $V=|S|h_i$, and $L=h_i$, as in \citet{1963MWRv...91...99S}. The trace-free shear tensor is used so that solid-body rotation and purely compressive or expanding flows do not contribute to the diffusion coefficient. The smoothing length is selected as the length scale because it is the smallest scale that can be resolved in an SPH simulation, and to be consistent with $|S|$, which is calculated using particles within the smoothing length.

In Paper \textsc{I} we found that metal-rich outflows are not produced if diffusion is too strong, because diffusion strips the metals from the outflowing gas before it escapes the disk. Hence we have selected one of the weaker diffusion models to correctly capture the observed metal-rich outflows. However, diffusion in this context is difficult to constrain through experiment or by comparison with observation.

In addition to the modifications of Paper \textsc{I}, we also include an external tidal field, which we describe in more detail in section~\ref{tidesec}.

We set the star formation efficiency to $\epsilon_*=0.02$ and the threshold density to $n_\mathrm{H}^{\phantom i}=1$ cm$^{-3}$. Prior to Paper \textsc{I}, we performed a series of tests with different values of the parameters in the star formation and feedback algorithm, and found the results were not sensitive to modest changes to these values, except for combinations of parameters that produced dramatic star formation bursts that destroyed the gas disk entirely.

Our simulation time is $2.5$ Gyr. This is sufficient time for the dwarf galaxies in each run to pass their peak star formation rate and to move onto a more quiescent phase (see section~\ref{starformsection}), and for these galaxies to complete a significant portion of their orbit (see section~\ref{orbitsubsec}), while still being completed within a reasonable computing time.

\subsection{Tidal Field}\label{tidesec}

We assume that the dwarf galaxy orbits inside the halo of a more massive host galaxy, and we model the tidal field of this host galaxy analytically. The host is assumed to have an NFW profile \citep{1996ApJ...462..563N} with a concentration of $c = 12$. The mass of the host galaxy ranges from $5\times10^{11}$ M$_\odot$ to $10^{13}$ M$_\odot$, and this mass sets the strength of the tidal field. The masses of the host halos, and the tidal strength ($|\mathrm{d}F/\mathrm{d}R|$), tidal radius ($r_\mathrm{J}$), and orbital period for the dwarf galaxy in these halos are given in Table~\ref{ictable}. We describe the orbits, which determine the tidal strength, tidal radius, and orbital period, in section~\ref{orbitsubsec}.

\begin{table}
\begin{tabular}{ccccccc}
\hline\hline
Name & $M_\mathrm{halo}$ & $|\mathrm{d}F/\mathrm{d}R|$ & $r_\mathrm{J}$ &  $T_\mathrm{orbit}$ \\
~& (M$_\odot$)  & (km s$^{-1}$ Gyr$^{-1}$ kpc$^{-1}$) & (kpc) & (Gyr) \\
\hline
A & - & - & - & -\\
B & $5\times10^{11}$ & 0.16 &22.3 & 5.3\\
C & $1\times10^{12}$ & 0.60 &14.9& 3.2\\
D & $5\times10^{12}$ & 1.94 & 10.4& 2.1\\
E & $1\times10^{13}$ & 6.58 & 6.9& 1.3\\
\hline
\end{tabular}
\caption{\label{ictable} \textup{
Summary of simulation parameters. $M_\mathrm{halo}$ is the mass of the external ``tidal'' halo potential, $|\mathrm{d}F/\mathrm{d}R|$ is the magnitude of the gradient of the halo force-field at the dwarf galaxy's centre of mass (i.e. the strength of the tides), $r_\mathrm{J}$ is the Jacobi (i.e. tidal) radius, and $T_\mathrm{orbit}$ is the orbital period of the dwarf galaxy.
}
}
\end{table}

Our simulations are performed in the frame of the centre of mass (CoM) of the dwarf galaxy. The acceleration of each particle from the external potential is thus equal to
\begin{equation}
\mathbf{a}(\mathbf{r}_i)=\mathbf{a'}(\mathbf{r}_i+\mathbf{r'}_\mathrm{CoM})-\mathbf{a'}(\mathbf{r'}_\mathrm{CoM}),
\end{equation}
where $\mathbf{r}_i$ is the position of particle $i$ in the dwarf galaxy's CoM frame, $\mathbf{r'}_\mathrm{CoM}$ is the position of the dwarf CoM in the host galaxy frame, $\mathbf{a'(\mathbf{r}')}$ is the acceleration from the external potential of point $\mathbf{r}'$ in the host galaxy frame, and $\mathbf{a}(\mathbf{r}_i)$ is the acceleration of particle $i$ in the dwarf CoM frame. We also integrate the position of the dwarf CoM frame within the galactic potential. For simplicity, we do not include dynamical friction or the gravitational force of the dwarf on the host galaxy.

\subsection{Galaxy Models}

We model our dwarf galaxies as a disk of gas and stars within a dark matter halo, as is common \citep[e.g.][]{2006MNRAS.369.1021M,2010MNRAS.405.1723S,2011ApJ...726...98K,2013MNRAS.436.1191T,2014MNRAS.438.1208K,2016ApJ...822...91W,2016arXiv160500650F}. Although observed dwarf galaxies are typically irregular or elliptical in shape (dIrrs and dEs), there is evidence that these morphologies are generated by interactions, in particular through tidal stirring and harassment of dwarf disks. This has been frequently demonstrated in simulations \citep[e.g.][]{1998ApJ...495..139M,2001ApJ...547L.123M,2001ApJ...559..754M,2005MNRAS.364..607M,2011MNRAS.415.1783B,2011ApJ...726...98K}, and could explain the morphology-density relation \citep[e.g.][]{1980ApJ...236..351D,1986ApJ...300...77G,1991AJ....101..765F}. In this work we investigate the effects of tidal fields on the evolution of dwarf galaxies, and so it is most appropriate to select an initial morphology that has not yet experienced the effects of a tidal field. That is, a dwarf disk that represents a precursor to a dIrr or dE galaxy. Furthermore, the simple morphology of a dwarf galaxy allows us to produce consistent equilibrium initial conditions, while dIrr initial conditions can be complex and potentially artificial.

We select an initial disk mass of $5\times10^8$ M$_\odot$. This is on the upper range of dwarf galaxy masses, and is intended to represent a precursor to a disk-like irregular or Magellanic type galaxy, such as the Magellanic Clouds, or galaxies such as NGC 55 and NGC 625 in the Sculptor group. We select a disk gas fraction of $f_g=0.5$, a value that is bracketed by the SMC and LMC gas mass fractions in the simulations of \citet{2012MNRAS.421.2109B}.

We select a scale height for the stellar disk of $100$ pc, and a scale length of $540$ pc. For the gas disk, we select a scale length of $860$ pc, and vertical distribution of gas is set by the criteria of hydrodynamic equilibrium and hence varies across the disk. This is somewhat compact for a dwarf galaxy of this mass, but local dwarfs show a large scatter in their scale-lengths \citep{2012AJ....144....4M}, and our initial scale-lengths are well within the observed distribution. We comment on the likely effects of different choices of scale-length in section~\ref{discsection}.

The disks have a low initial metallicity ($[\alpha/\mathrm{H}]=-2$ for all $\alpha$ species, $[\mathrm{Fe}/\mathrm{H}]=-3$, giving $[\alpha/\mathrm{Fe}]=1$). Solar metallicities are taken from Table~2 of \citet{1995ApJS..101..181W}, and are only used in the initial conditions and to normalize the results we present. We emphasize that because the initial metallicity is low, the abundance ratios in our simulations are dominated by modelled star formation, and hence our results should not be very sensitive to different choices of solar metallicities. All particles have the same initial metallicity, and so any metallicity gradient can only result from subsequent evolution.

The disk consists of $5\times10^5$ particles, giving a mass resolution of $10^3$ M$_\odot$ per particle, where gas and star particles have the same mass. The disk is embedded within a dark matter halo of mass $9.5\times10^9$ M$_\odot$, consisting of $9.5\times10^5$ particles. The halo follows a NFW profile with a concentration of $c=10$.

\subsection{Orbits}\label{orbitsubsec}

Cosmological simulations suggest that satellite galaxies tend to have highly eccentric orbits \citep[e.g.][]{1997MNRAS.290..411T,2008ApJ...688..757H,2011MNRAS.412...49W,2015MNRAS.448.1674J} and that circular orbits are unlikely to be common, while proper motion studies of Milky Way dwarfs have found that several dwarfs appear to have moderate or low eccentricities \citep[e.g.][]{2006AJ....131.1445P,2007AJ....133..818P,2011ApJ...741..100L}, although the uncertainties are often large. Here, we use circular orbits to provide a continuous and smooth tidal field, to examine the long-term effects of a tidal field on a dwarf galaxy, without needing to explore the vast parameter space of interactions between galaxies in an explicitly modelled cosmological environment. The effects of tides should be sensitive to orbital parameters, but a full exploration of this parameter space is beyond the scope of this work. Tidal effects become stronger with decreasing eccentricity \citep[e.g.][]{2010MNRAS.405.1723S,2015MNRAS.454.2502S} at constant periapsis, because the satellite galaxy spends more time near its periapsis. Hence our models represent the greatest tidal effects expected from our modelled host halos on a bound dwarf galaxy at our chosen periapsis of $100$ kpc.

In runs B-E, the dwarf galaxy follows a near-circular orbit through the host potential well. Run A does not include a tidal field, and the dwarf is stationary. In runs B-E, the dwarf is initially oriented face-on towards the centre of the potential well, and remains at a near-constant distance from the centre of the halo. As the dwarf orbits the host halo, its absolute orientation does not change, and so the dwarf naturally alternates between face-on and edge-on orientations with respect to the host. The orbital speed of the dwarf galaxy depends on the tidal potential, ranging from $115$ km\,s$^{-1}$ for B up to $460$ km\,s$^{-1}$ for E. These produce orbital periods from $1.3-5.3$ Gyr as shown in Table~\ref{ictable} -- i.e. when tides are stronger, they are also varying more rapidly. 

These orbits gives Jacobi radii that are large compared with the scale lengths of the dwarf galaxies -- i.e. these are all fairly weak tides. The intention here is to investigate how mild tides can affect the chemodynamical evolution of the dwarf galaxy, rather than the destruction of dwarf galaxies by very strong tides. The very strongest tides will also likely occur in closer to the host galaxy where ram pressure is stronger, and we have not included ram pressure in these simulations.

\subsection{Definitions}\label{nomensec}

In this paper, we use a coordinate system where $z$ is the angular momentum axis of the dwarf, and $x=y=z=0$ are cartesian coordinates for the centre of the dwarf. We also use the cylindrical radius $R=\sqrt{x^2+y^2}$. In our analysis, we have divided the domain of the simulation into different zones. With $R$ and $z$ in kpc units, these zones are defined as follows:

\begin{enumerate}
\item The inner disk zone, $0\le R<2$, $|z|<2$.\\
\item The intermediate disk zone, $2\le R<4$, $|z|<2$.\\
\item The outer disk zone, $4\le R<6$, $|z|<2$.\\
\item The vertical outflow zone, $0\le R<6$, $2\le |z|<5$.\\
\end{enumerate}

In our analysis, we calculate the scale length of the simulated disks. We do this by producing a radial cumulative mass profile $M(R)$ for the gas disk at $t=2.5$ Gyr, and fitting $M(R)$ across the inner $20$ kpc with the equation 
\begin{equation}
M(R) = -Ah_r\mathrm{e}^{-R/h_r}(2h_r^2+2Rh_r+R^2)+B,
\end{equation}
where $h_r$ is the radial scale length of the gas disk, and $A$, $B$, and $h_r$ are fitted with a standard least-squares fitting method. This profile is the equivalent to a disk with an exponential surface-density profile with a scale length of $h_r$, but fitting the cumulative mass profile reduces some of the numerical error. The fitting is evenly weighted at all radii, and hence is dominated by the outer low-density regions of the disk rather than by the small inner disk zone. Therefore, this scale length is a measure of how rapidly the density drops at large radii, and can be used as a measure of the truncation of the outer disk from tidal effects.

\makeatletter{}\section{Results}\label{resultssection}
\makeatletter{}
\begin{figure*}
\begin{center}
\includegraphics[width=\textwidth]{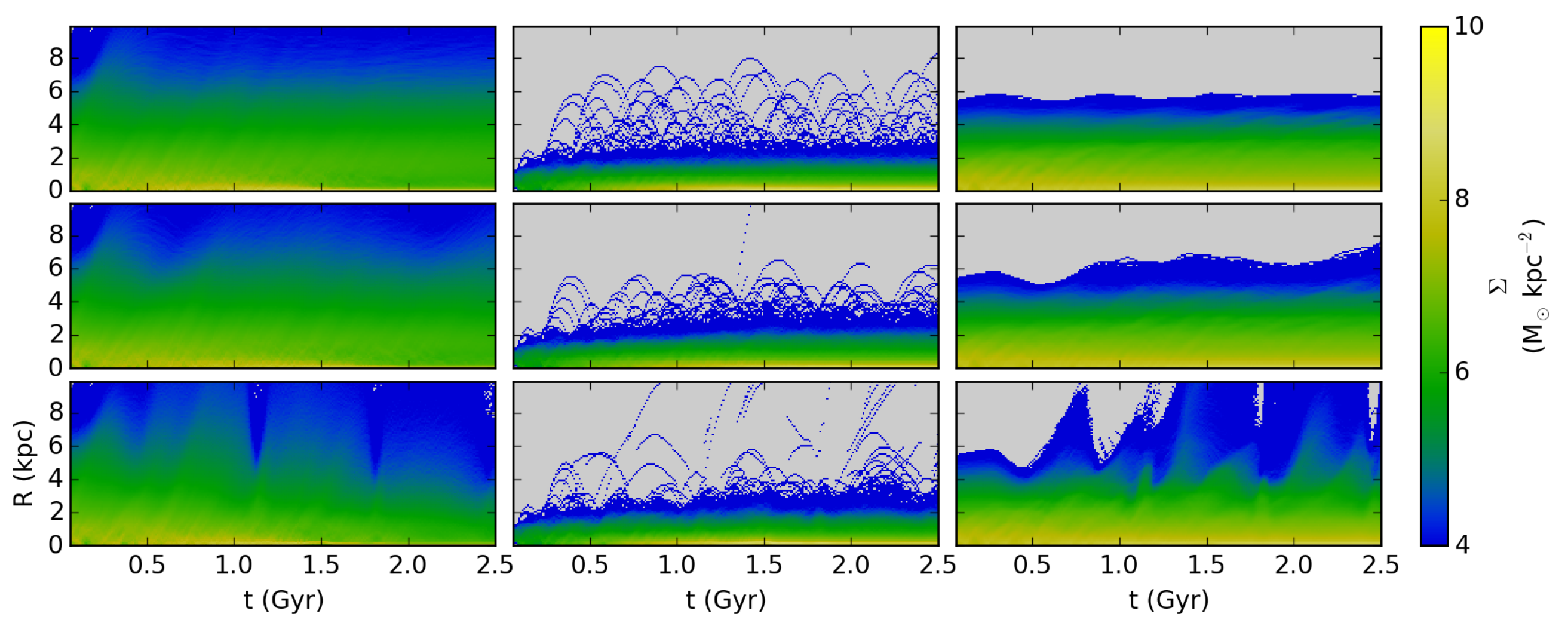}
\end{center}
\caption{\label{rothistsurf}
Evolution of radial profiles of surface density, for gas (left column), formed stars (center column), and initial stars (right column), for runs A, C, and E, ordered from top to bottom.
}
\end{figure*}

\begin{figure*}
\begin{center}
\includegraphics[width=\textwidth]{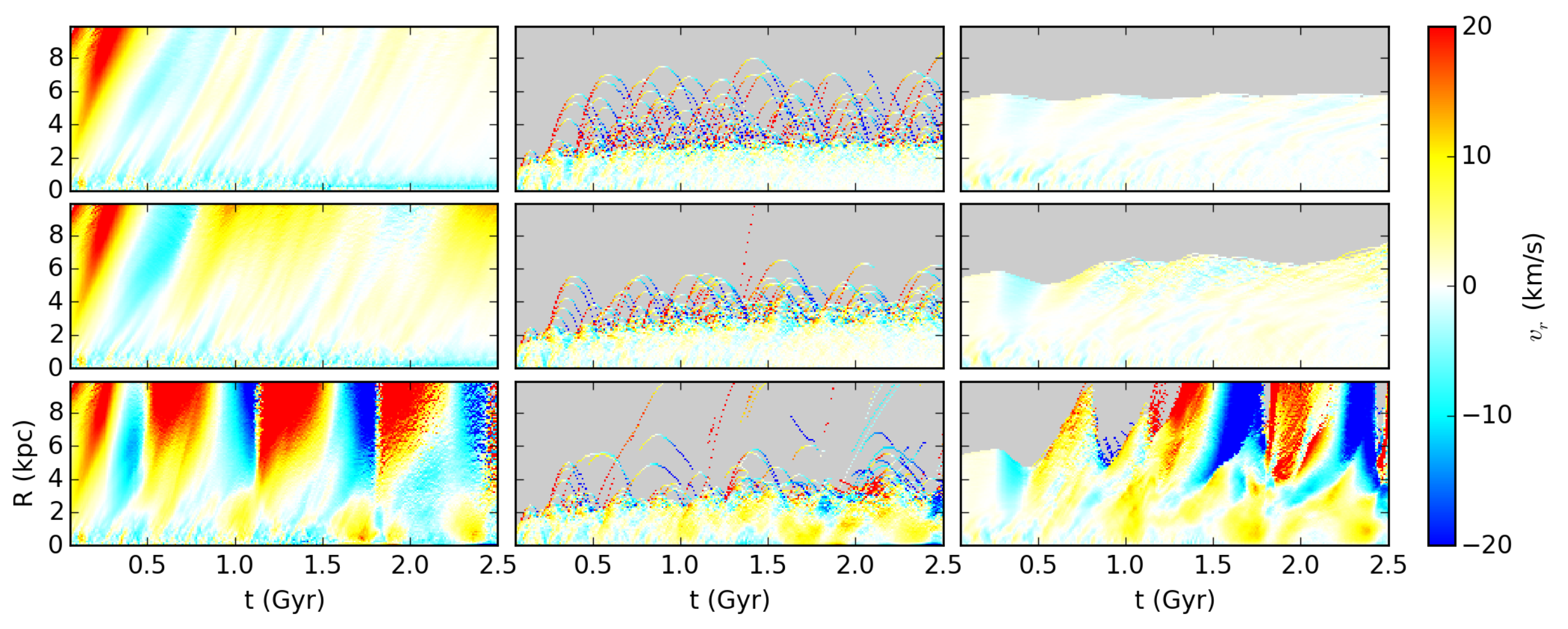}
\end{center}
\caption{\label{rothistvr}
Evolution of radial profiles of radial velocity, for gas (left column), formed stars (center column), and initial stars (right column), for runs A, C, and E, ordered from top to bottom.
}
\end{figure*}

\begin{figure*}
\begin{center}
\includegraphics[width=\textwidth]{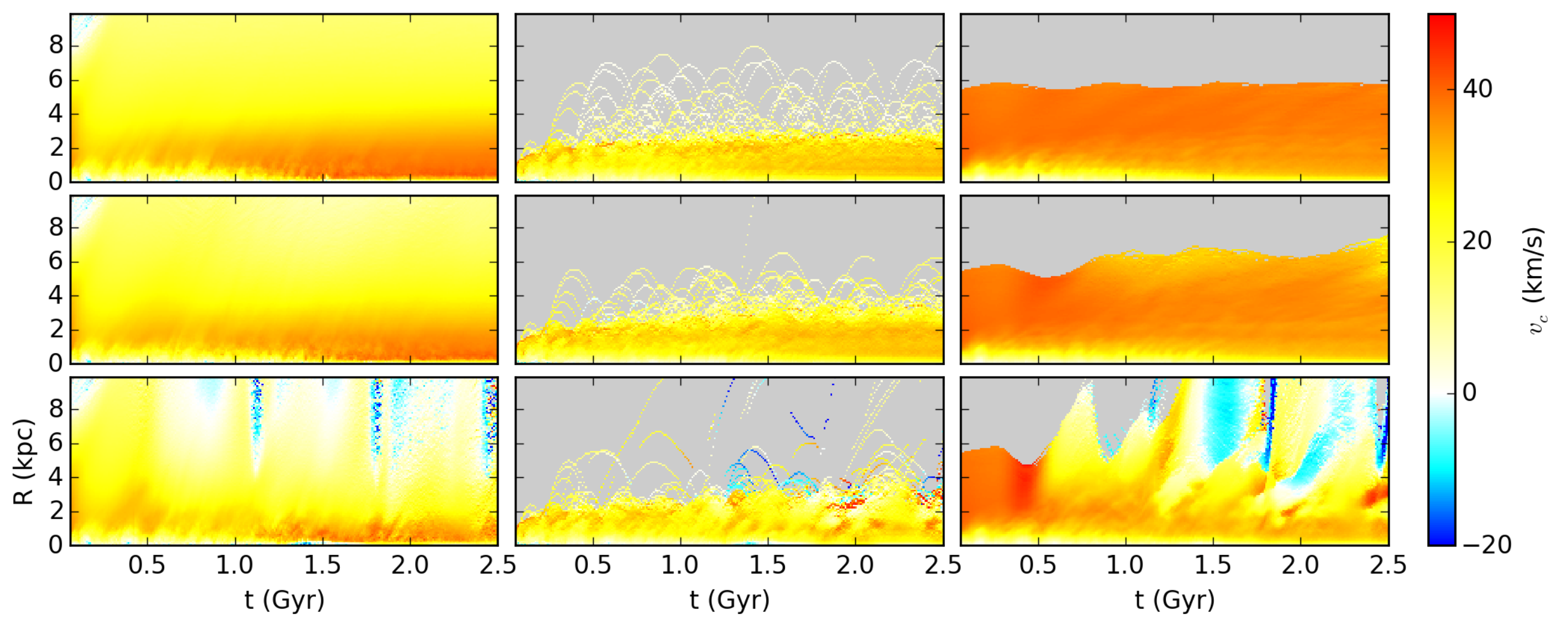}
\end{center}
\caption{\label{rothistvc}
Evolution of radial profiles of circular velocity, for gas (left column), formed stars (center column), and initial stars (right column), for runs A, C, and E, ordered from top to bottom.
}
\end{figure*}

\begin{figure*}
\minipage{0.66\textwidth}
\includegraphics[width=\linewidth]{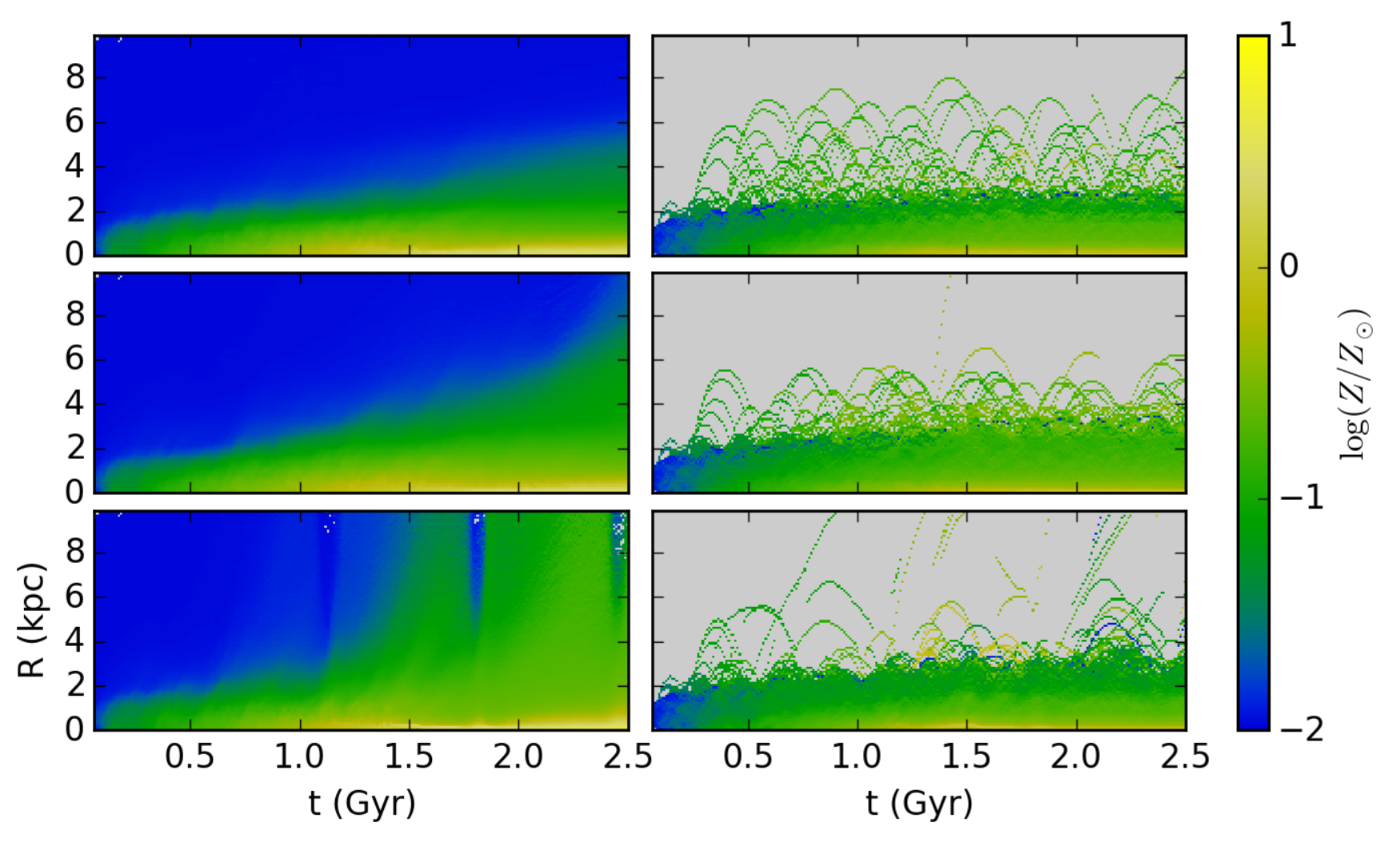}
  \caption{Evolution of radial profiles of metallicity, for gas (left column), and formed stars (right column) for runs A, C, and E, ordered from top to bottom}\label{rothistZ}
\endminipage\hfill
\minipage{0.33\textwidth}
\includegraphics[width=\linewidth]{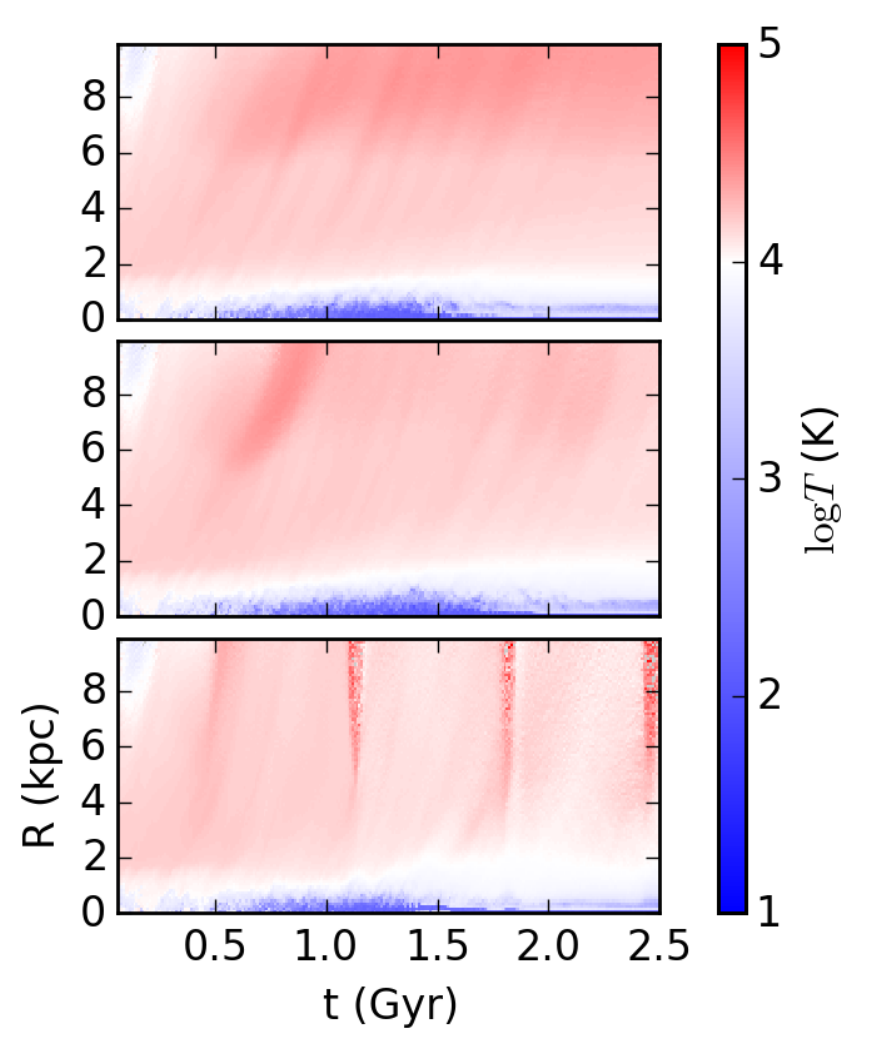}
  \caption{Evolution of radial profiles of gas temperature, for runs A, C, and E, ordered from top to bottom}\label{rothistT}
\endminipage\hfill
\end{figure*}

\begin{figure*}
\begin{center}
\includegraphics[width=\textwidth]{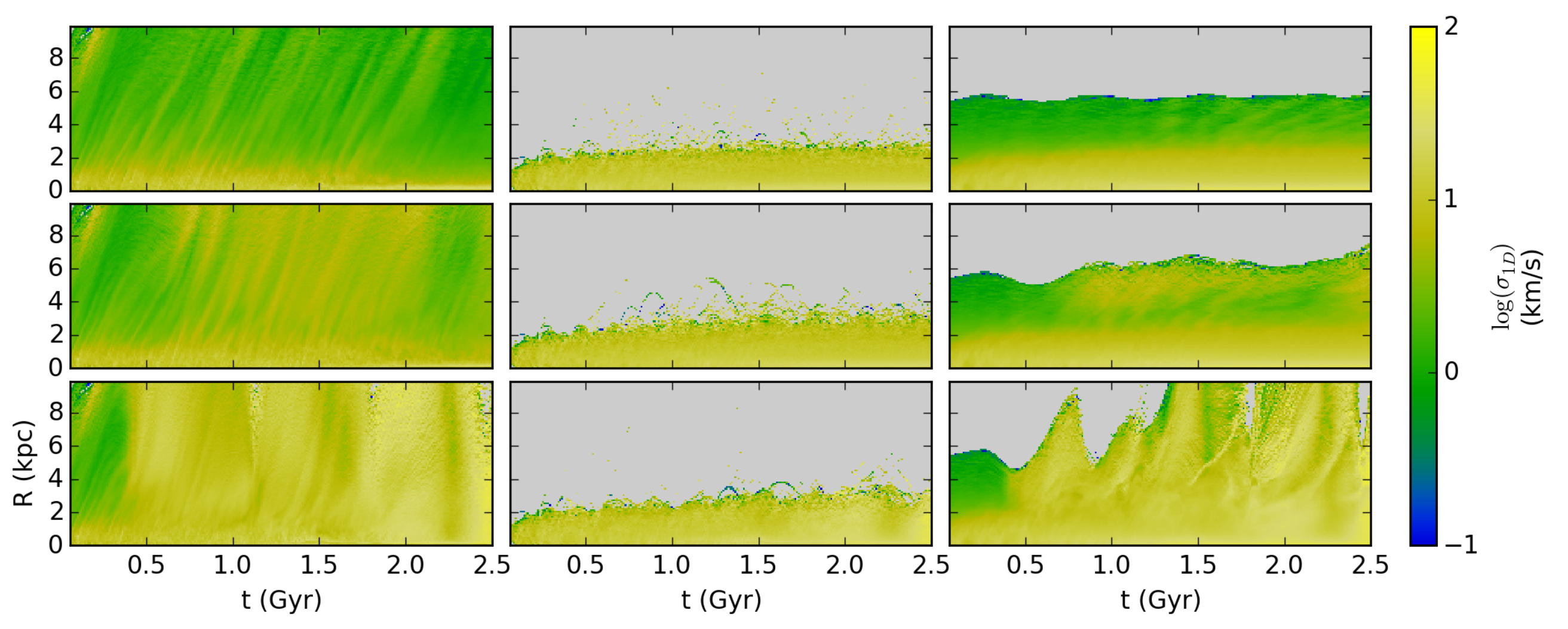}
\end{center}
\caption{\label{rothistdv}
Evolution of radial profiles of radial velocity dispersion, for gas (left column), formed stars (center column), and initial stars (right column), for runs A, C, and E, ordered from top to bottom.
}
\end{figure*}

\begin{figure*}
\includegraphics[width=.98\textwidth]{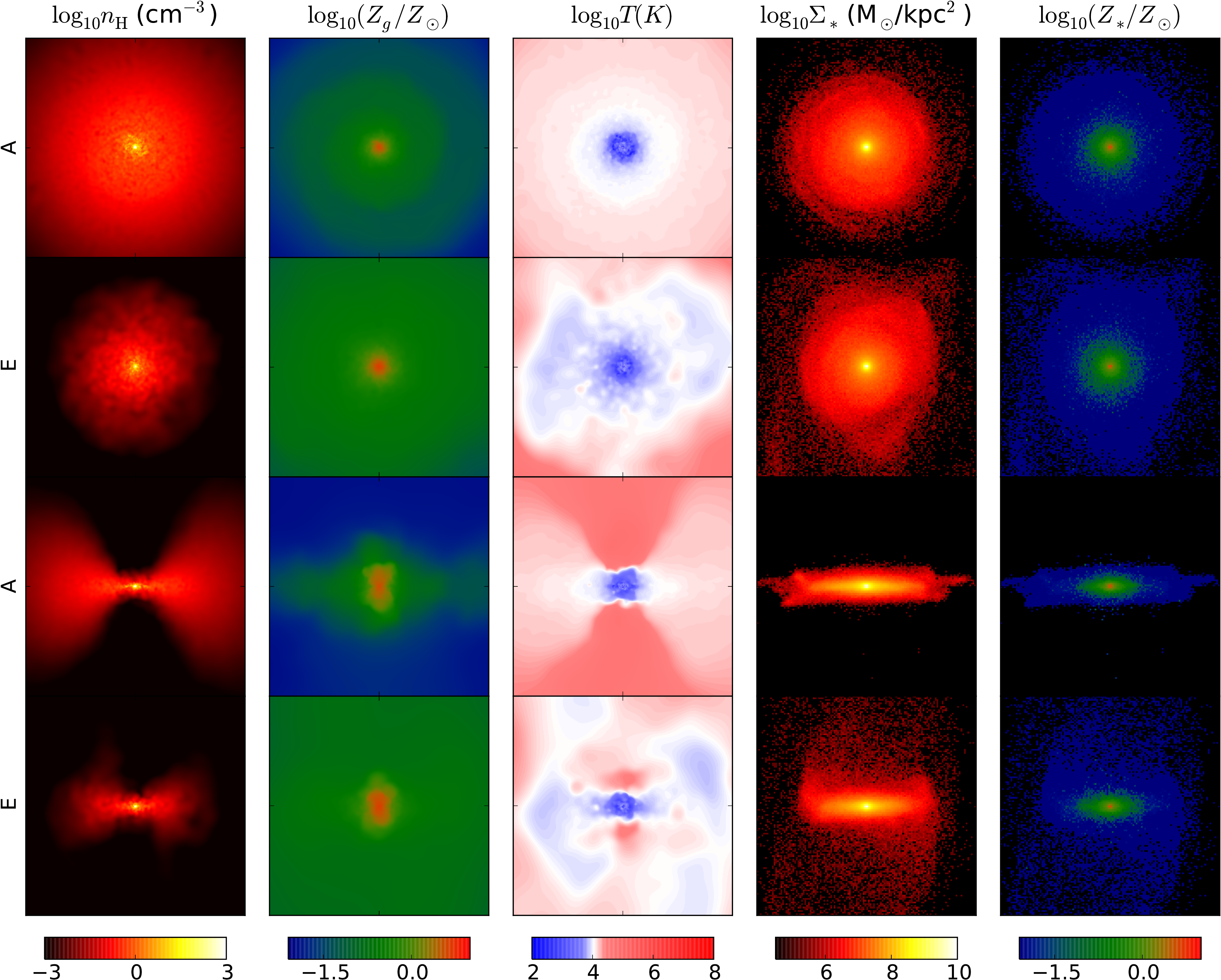}
\caption{\label{face2G5yr}
Face-on (top two rows) and edge-on (bottom two rows) slices of runs A and E at $t=2.5$ Gyr. Each box has dimensions of $10\mathrm{~kpc}\times10$ kpc. Here $\Sigma_*$ and $Z_*$ describe the entire population of stars (both formed and initial stars).
}
\end{figure*}

\subsection{General Evolution}\label{genev}

To summarize the evolution of our simulated galaxies, we produce heat-maps showing the evolution over $2.5$ Gyr of radial profiles of the important quantities for the different species of particle. We calculate radial profiles at each output dump for three different species of particle: gas, stars formed within the simulation, and stars present in the initial conditions. The significance of plotting the initial stars separately is that they act as tracers of purely dynamical and gravitational processes, while the properties of stars formed within the simulation can be largely determined by the hydrodynamics of the gas that formed them. The calculated values are temperature (only for gas), metallicity (except for initial stars), circular velocity, radial velocity, radial velocity dispersion, and surface density. We do not calculate the metallicity profiles for the initial stars, because their metallicities are a constant value set by the initial conditions

The radial profiles across the first $2.5$ Gyr of the simulations are plotted in Figures~\ref{rothistsurf}-\ref{rothistdv}. We plot simulations A, C, and E as representative examples. To calculate these radial profiles, we divide the disk into bins by radius in increments of $100$ pc, only including particles within $2$ kpc of the disk plane. For the radial velocity, circular velocity, metallicity, and temperature profiles, we plot the median value in each bin. We use the median instead of the mean because it is less affected by outlying particles with extreme values. The surface density and radial velocity dispersion are ensemble properties, and are calculated using all particles of the appropriate species in the bin.

First, we examine run A -- the top rows of Figures~\ref{rothistsurf}-\ref{rothistdv}. After an initial period of equilibration, the radial surface density profile (Figures~\ref{rothistsurf}) settles down across most of the disk. However, outflows and inflows are still present in the radial velocity profiles (Figure~\ref{rothistvr}), particularly in the gas component, indicating that the origin is from feedback processes. Feedback is fairly dramatic here, and drives large-scale outflows of gas, which then fall back towards the center of the galaxy. These outflows are faintly visible in Figure~\ref{rothistsurf} as streaks that move to greater radii over time.

More clearly, the gas in the central $2$ kpc becomes increasingly concentrated. This causes the circular velocities in the central $2$ kpc to increase (Figure~\ref{rothistvc}) as the mass concentration causes the rotation curve to locally become more Keplerian. We further discuss the causes and consequences of this concentration of gas in the center in section~\ref{starformsection}.

A metallicity gradient also develops in the gas (Figure~\ref{rothistZ}). Gas is enriched in the central regions by star formation, then spreads outwards through feedback-driven flows and diffusion. This star-forming gas is visible in the temperature plot (Figure~\ref{rothistT}) as the cool ($T\lesssim10^3$ K) gas in the center of the galaxy. Feedback stirs the gas, producing large velocity dispersions in the center (Figure~\ref{rothistdv}), and producing hot radial outflows, visible as streaks of warm gas moving to greater radius over time. This produces a temperature gradient with a positive slope -- cool star-forming gas in the center, and warm gas in the outer regions.

Next, we examine the simulations containing an external tidal potential (all other rows of Figures~\ref{rothistsurf}-\ref{rothistdv}). Tides pull the gas and stars outwards when the galaxy is edge-on to the center of the external potential, and this material falls back inwards when the galaxy is face-on to the potential, giving a periodic process with a period approximately equal to half of the orbital period. This effect is strongest and most rapid in run E, where it is clearly visible in the surface density (Figure~\ref{rothistsurf}) and radial velocity (Figure~\ref{rothistvr}) plots. In runs B-E, the stars in the outer disk ($R>4$ kpc) also develop large radial velocities with strong tides. Here however, the surface density of the outer stellar disk is low, and in most cases the majority of stars remain in the inner regions of the disk where they are not strongly affected by the tidal field. Only Run E shows significant radial velocities in the stars in the central region.

The effect of tides on the metallicity distribution is particularly striking (Figure~\ref{rothistZ}). The tidally induced outflows of gas carry enriched material to greater radii as the strengths of tides increases. Some oscillation is also visible as the disk changes orientation with respect to the gradient of the external potential. We discuss tidal stripping in greater detail in section~\ref{stripsection}, and tidal effects on the metallicity gradient in section~\ref{flatsection}.

Tidal stirring increases the velocity dispersion of the disks (Figure~\ref{rothistdv}), even in runs C where tides are weak. Run E becomes extremely disturbed at late times, with velocity dispersions reaching as high as $100$ km/s in run E. These velocity dispersions have consequences for the stability and star formation rate of the disks, which we discuss in section~\ref{starformsection}.

To illustrate the late-time states of the simulated galaxies in greater detail, Figure~\ref{face2G5yr} shows slices through the $z=0$ pc and $x=0$ pc planes (i.e. face-on and edge-on views) for runs A and E at $t=2.5$ Gyr. In all simulations, the gas takes the form of a cool/warm flared disk ($T\le10^4$ K) within a low density hot halo, with some spiral structures and hot feedback bubbles visible. Vertical and horizontal metallicity gradients are also present.

The low-density outer regions are noticeably affected by tides, and this is most clearly visible in the temperature and gas metallicity plots. Without tides (run A), the warm ($T\sim10^4$ K) disk gas takes the form of a simple flared disk, but with strong tides (run E), this outer gas is greatly distorted. The stronger tides also show metallicity that is more vertically distributed.

The effects of tides on the distribution of stars is weaker, as the stellar population is not directly affected by feedback events, and star particles remain deep within the dwarf's potential well. Nevertheless, with strong tides, a warp is visible in outer regions of the stellar disk. The face-on plots also indicate there may be some asymmetry in the stellar distribution when strong tides are present.

\makeatletter{}\subsection{Star formation and gravitational instabilities}\label{starformsection}

\begin{figure}
\begin{center}
\includegraphics[width=.98\columnwidth]{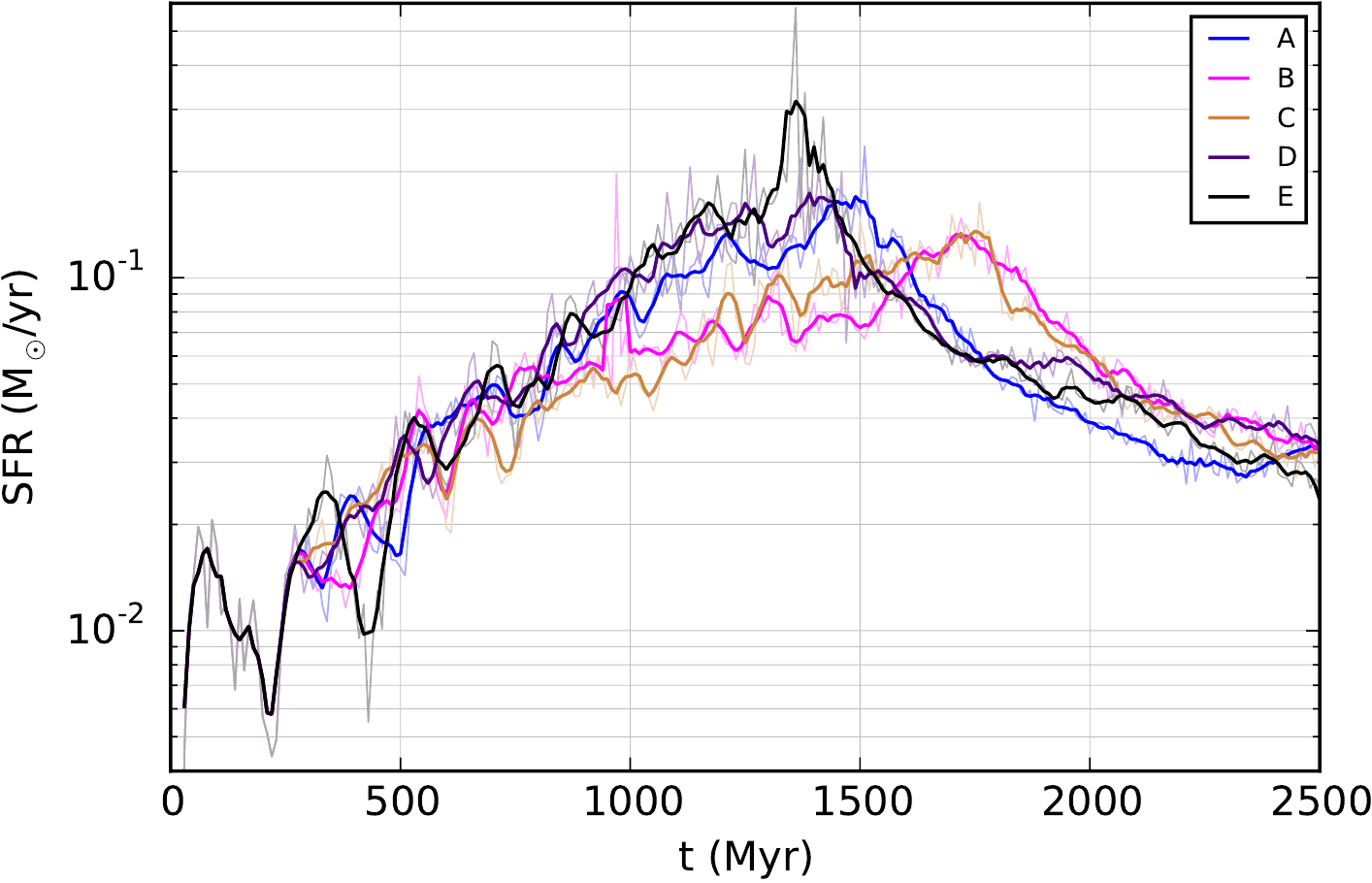}
\end{center}
\caption{\label{SFR}
Global star formation rates. The thin lines are the instantaneous star formation rates, the thick lines are the smoothed values.
}
\end{figure}

\begin{figure}
\begin{center}
\includegraphics[width=.98\columnwidth]{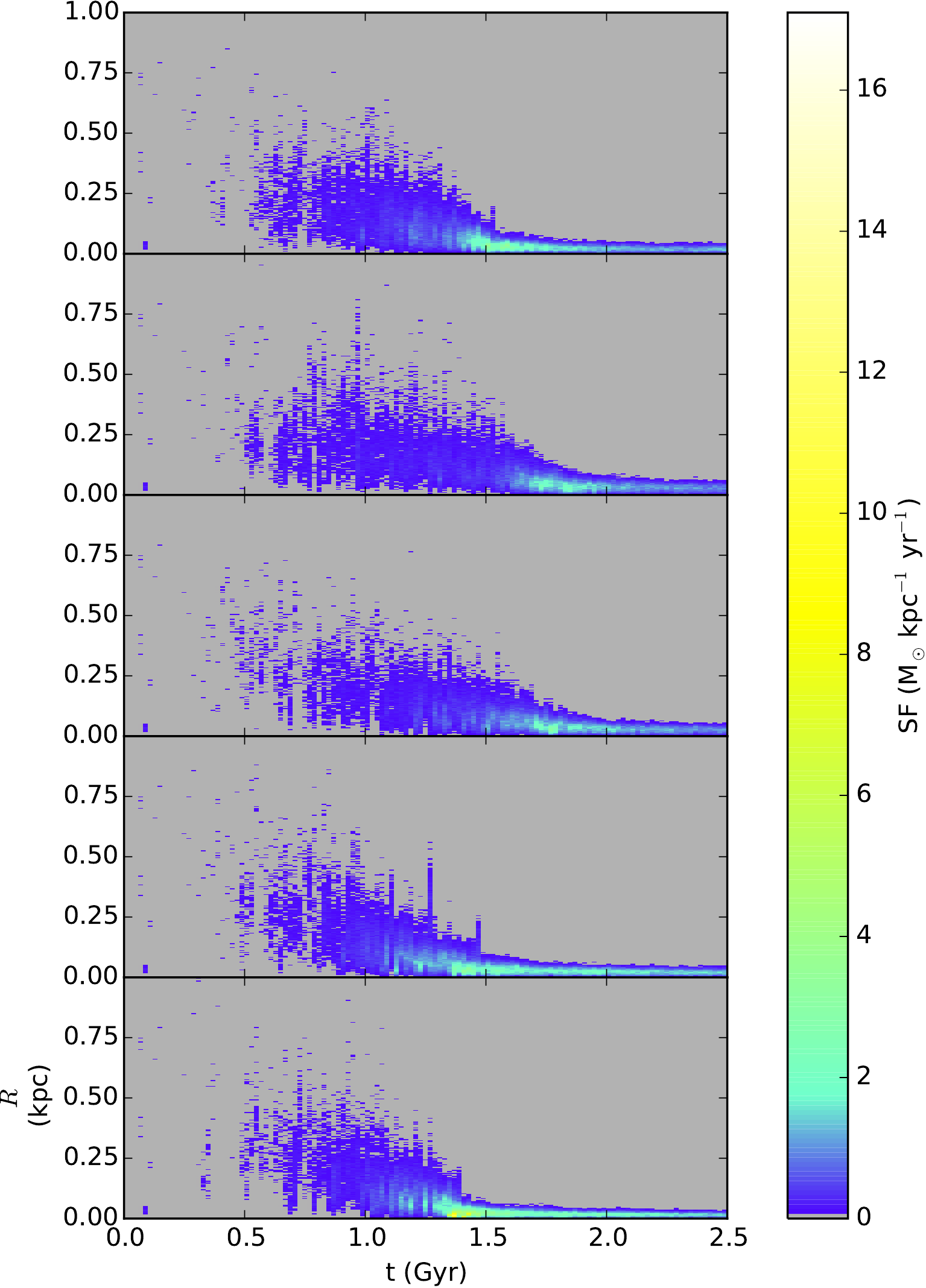}
\end{center}
\caption{\label{sfpos}
Positions of star formation events for runs A-E, ordered from top to bottom.
}
\end{figure}

\begin{figure}
\begin{center}
\includegraphics[width=.98\columnwidth]{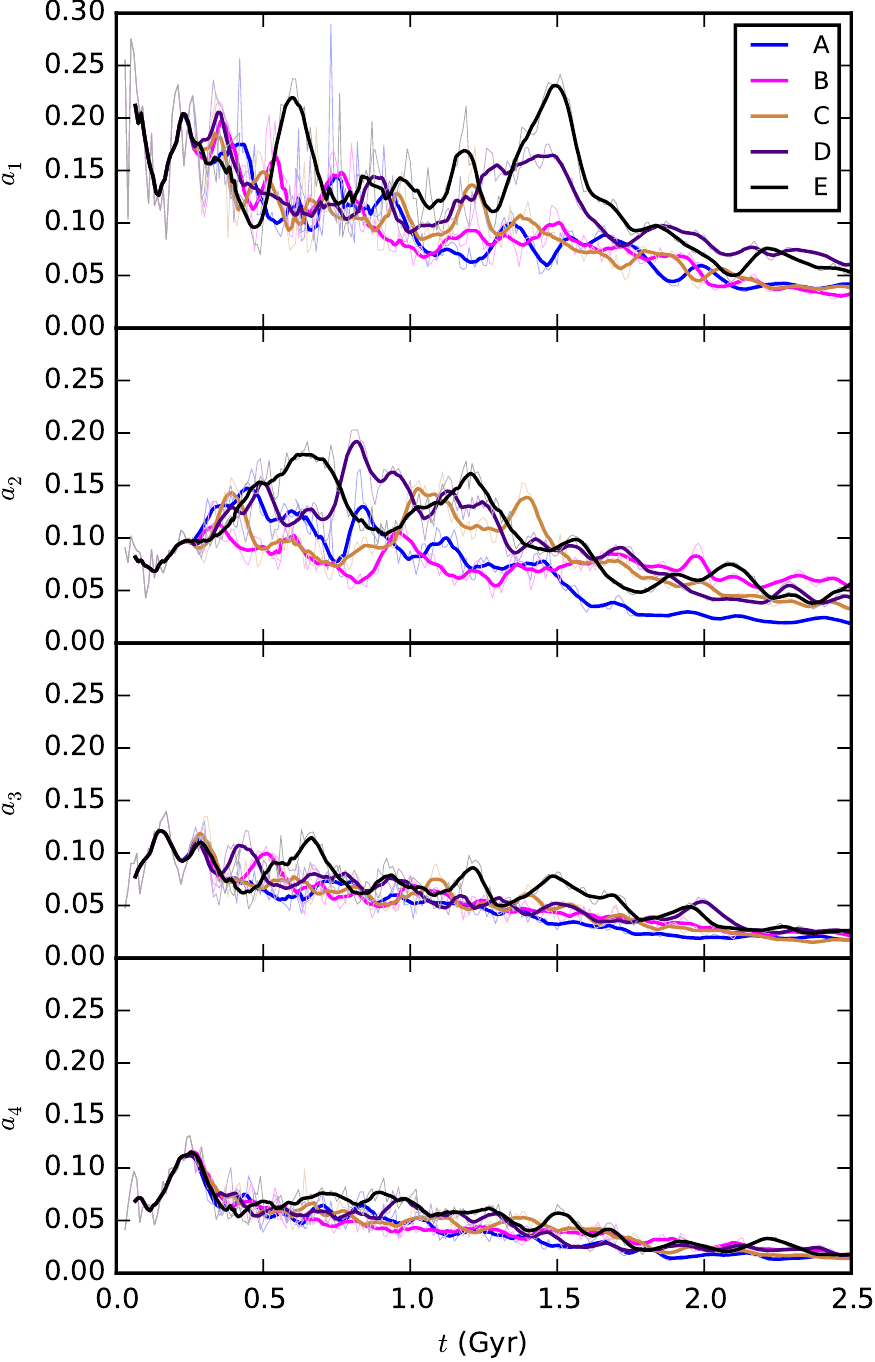}
\end{center}
\caption{\label{barstr}
Strength of gravitational instabilities, $a_1$-$a_4$. The faint lines show the values at each output dump, while the thick lines are smoothed values.
}
\end{figure}

\begin{figure}
\begin{center}
\includegraphics[width=.98\columnwidth]{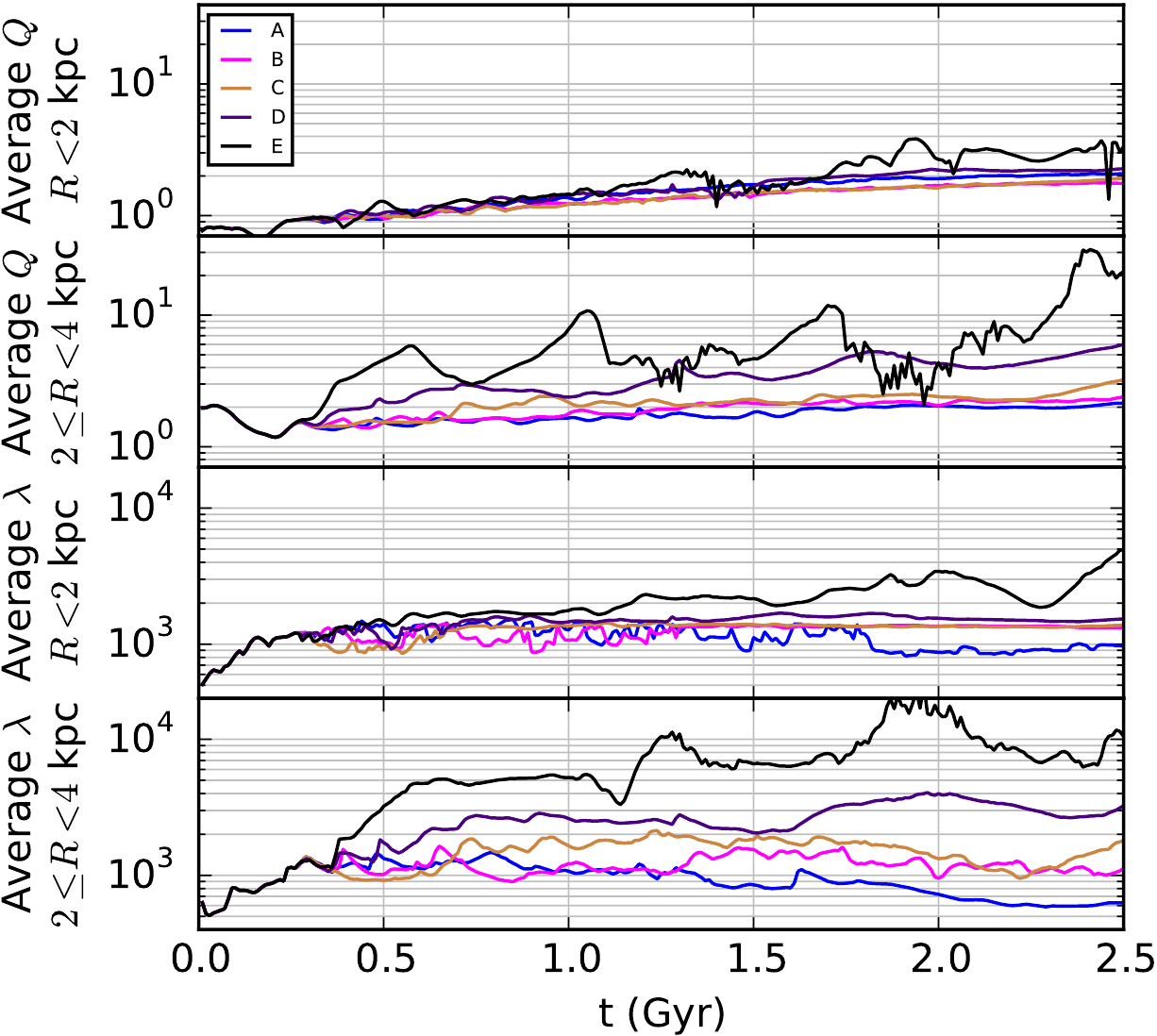}
\end{center}
\caption{\label{qtime}
Evolution of area-weighted geometric mean three-component $Q$ and $\lambda$ (in units of pc) for the inner disk and intermediate disk zones.
}
\end{figure}

\begin{table}
\begin{tabular}{ccccccc}
\hline\hline

Name & $|\mathrm{d}a/\mathrm{d}R|$ & $M_\mathrm{SF}$ & $h_r$ & $\Delta M_Z$ & $\langle a_2 \rangle$ & $\langle Q \rangle$ \\
\hline
A	&	$0.$	&	$145$	&	1.00	&	4.03	&	0.110	&	0.921\\
B	&	0.16	&	$137$	&	0.96	&	3.80	&	0.090	&	0.911\\
C	&	0.60	&	$136$	&	0.98	&	3.86	&	0.095	&	0.912\\
D	&	1.94	&	$160$	&	0.82	&	4.58	&	0.109	&	0.937\\
E	&	6.58	&	$166$	&	0.62	&	4.68	&	0.124	&	0.946\\
\hline
\end{tabular}
\caption{\label{qsfr} \textup{
Star formation, tides, and instability parameters for all runs. From left-to-right the columns are: the name of the run; $|\mathrm{d}a/\mathrm{d}R|$ -- the strength of the tidal force measured by the magnitude of the gradient of the acceleration field at disk center in units of km s$^{-1}$ Gyr$^{-1}$ kpc$^{-1}$; $M_\mathrm{SF}$ -- the total mass of star formation by $t=2.5$ Gyr in units of $10^6$ M$_\odot$; $h_r$ -- the scale length of the gas disk in units of kpc; $\Delta M_Z$ -- the total mass of metals formed by star formation by $t=2.5$ Gyr in units of $10^6$ M$_\odot$; $\langle a_2 \rangle$ -- the mean value of $a_2$ over the first $0.75$ Gyr; and $\langle Q \rangle$ -- the mean value of the $Q$ for the inner disk region over the first $0.75$ Gyr.
}
}
\end{table}

Figure~\ref{SFR} shows the star formation rates as a function of time for all five runs presented here. In all runs, as gas cools and gravitational instabilities develop, the star formation rate gradually increases, before reaching a peak and decaying away as gas is consumed or ejected. These peaks occur at different times, and the earlier the peak is, the higher its value. The peak star formation rates are large, but within the range of observed starbursting dwarfs \citep{1988ApJ...334..665F}.
B and C produce peak star formation rates lower than that of A, while E has a significantly larger peak star formation rate. The gentle stirring of the weak tides may act to stabilize the disk against star formation, but this effect may not be large enough to be significant. However, the dramatic impact of the strong tides in run E amplifies the star forming instabilities, producing a strong and early burst of star formation.

The radial positions of star formation events over time are plotted in Figure~\ref{sfpos}. The clear general trend in all five models is that the star forming region gradually moves inwards over time. Centrally concentrated star formation in dwarfs has been noted in observations \citep{2011MNRAS.417.1643K}, particularly in blue compact dwarfs (BCDs) of the nE type (i.e. dwarf galaxies with a clearly defined nucleus) as noted in \citet{1986sfdg.conf.....K} and \citet{1996A&AS..120..207P}. In our simulations, rapid central star formation occurs once the gas has become sufficiently concentrated (Figure~\ref{rothistsurf}), providing a large reservoir of cool gas in the center of the galaxy (Figure~\ref{rothistT}). The star formation is therefore primarily the result of gas that has been driven inward and become cool and dense, and the star formation rate is determined primarily by the rate of gas inflow. This rate is affected by the presence of tides -- gas inflow is accelerated by strong tides, and may perhaps be slowed by weak tides. This trend is summarized in Table~\ref{qsfr}. Both the total mass of star formation, and the total mass of metals formed by $t=2.5$ Gyr greatly increases with strong tides (D \& E), indicating that star formation is partially triggered by the presence of tides. There is also a small decrease in the total star formation mass in the runs with weak tides (B \& C). The radial scale length also decreases with increasing tide strength, but this is the result of tidal stripping truncating the gas disk, and not of gas becoming centrally concentrated by gravitational instabilities. We discuss this in more detail in section~\ref{flatsection}.

Careful analysis is required to determine the cause of this process. Gravitational instabilities can produce non-axisymmetric features such as bars and spiral arms that transport gas inwards. Feedback can induce a large velocity dispersion in the gas, generating an effective viscosity that may also transport gas inwards. The interactions between these effects can be complex and difficult to predict. Indeed, it has been found that feedback can somewhat {\em reduce} the inflow rate by smoothing and heating the gas, stabilizing it against gravitational instabilities \citep{2015ApJ...814..131G,2016arXiv160500646G}. We investigate the role of gravitational instabilities by measuring the amplitude of non-axisymmetric modes, and by performing a Toomre-like stability analysis.

To measure the strength of non-axisymmetric instabilities, we perform a Fourier analysis based on the method of \citet{2013MNRAS.429.1949A}. We divide the inner $5$ kpc of the disk into 50 radial bins, and calculate the magnitude $a_m$ of the Fourier modes for $m=1,2,3,4$, in each radial bin. All gas and star particles within $2$ kpc of the disk plane are included in this analysis. At each time we select the maximum values of $a_m$ across all radial bins.

This process can be sensitive to the definition of the center of the galaxy, causing even modes to be identified as odd modes or vice versa. This does not have a large impact on our analysis, as we are chiefly concerned with the strength of gravitational instabilities in general, and not with the relative strengths of different modes, as even lopsided modes can drive gas inwards \citep{2009PhR...471...75J}. Nevertheless, we attempt to reduce any error in determining the center of the dwarf galaxy. We find that a simple center-of-mass calculation was not sufficiently accurate, because the external tidal field does not conserve momentum. Particles at large distances from the center of the dwarf can experience particularly strong tidal forces, and are also heavily weighted in a center-of-mass calculation. To reduce this error, we set a minimum density threshold for inclusion in the calculation of the galaxy center. A sum is then performed of the position of all included particles, weighted according to their mass and the log of their density. In GCD+, the local hydrodynamic density is calculated for star particles, and so both gas and star particles can be included in this weighted sum. Dark matter particles are not included. The full sum across the mass and position of all star and gas particles ($m_i$ and $\mathbf{r}_i$) weighted according to their density $\rho_i$ to calculate the center $\mathbf{r}_c$, proceeds as follows:

\begin{align}
W_i = &
\begin{cases}
\log_{10}(\rho_i/\rho_0) & \textrm{if}\ \rho_i>\rho_0,\\
0. & \textrm{otherwise}
\end{cases}\\
\mathbf{r}_c = & \left(\sum_i W_i m_i \mathbf{r}_i \right) \Big/ \sum_i m_iW_i
\end{align}where $\rho_0$ is the density threshold. Too small a value of $\rho_0$ will give a large weighting to distant particles, and too large a value of $\rho_0$ will give a center that varies noisily with time, based on the positions of a small number dense clumps. From a visual inspection of the results, we found that a value of $\rho_0=10^{-25}$ g~cm$^{-3}$ appears to avoid both of these extremes for our galaxies, but we note that in general the best choice for $\rho_0$ will depend on the numerical and physical details of the particular simulation.

The resulting strengths of the Fourier modes are plotted in Figure~\ref{barstr}. In all runs, $a_3$ and $a_4$ are small throughout the simulation, quickly ($t<0.25$ Gyr) reaching an initial peak and then decreasing as gas is consumed and driven inwards. There is no clear indication of a relationship with tide strength from these plots alone.

The strengths of lopsided modes ($a_1$), and bar or two-armed spiral modes ($a_2$) also generally show a long decay, but with a larger maximum amplitude, and with greater variation between the runs. It appears that $a_1$ and $a_2$ are both generally larger in the runs with the strongest tides, and that suggesting that lopsided $m=1$ modes are more significant than $m=2$ modes, although we note again that the relative strengths of these modes is sensitive to small changes in the definition of the center of the galaxy.

To investigate if gravitational instabilities at early times are responsible for driving gas inwards and inducing centrally concentrated star formation, we have tabulated the mean value of $a_2$ in the first $0.75$ Gyr of each simulation against the total mass of stars formed in that simulation in Table~\ref{qsfr}. A general trend is visible, where the simulations with the strongest tides and the strongest bars also have the strongest star formation, which suggests a potential connection. We find that $a_1$, $a_3$, and $a_4$ show a similar correlation. 
The Fourier amplitudes after $t=0.75$ Gyr do not appear to be related to the star formation, because the gas is already well on the way to becoming centrally concentrated. At late times ($t>1.5$ Gyr), the $m=2$ mode becomes weak in all simulations, although the runs with a tidal field (B-E) consistently have a greater bar strength than A, suggesting that the perturbations of the external potential are helping to drive the instability. If the measured Fourier amplitudes were not describing ``true'' bars or spirals, but instead were a result of the the tidal field directly stretching the disk along the radial axis of the host halo, then we would expect $a_2$ at late times ($t>1.5$ Gyr) to increase with the strength of the tidal fields, and no such trend is observed.

To clarify the role of these instabilities, we perform a Toomre-like stability analysis by calculating the mutli-component Q stability parameter $Q_\mathrm{RF}^{\phantom i}$, and the characteristic instability wavelength $\lambda_\mathrm{RF}^{\phantom i}$, following the recipe given by \citet{2013MNRAS.433.1389R}, as follows:

First, we calculate $Q$ for the stellar and gas components. We calculate $Q$ separately for the initial stars and formed stars, because these are distinct dynamical components, as shown in Figures~\ref{rothistsurf}-\ref{rothistdv}. This gives three separate species -- initial stars, formed stars, and gas. The $Q$ parameters are calculated through
\begin{equation}
Q_i = \frac{\kappa \sigma_i}{\pi G \Sigma_i},\\
\end{equation}where $i$ denotes each of the three components (initial stars, formed stars, and gas), $\kappa$ is the epicyclic frequency, $\sigma$ is the radial velocity dispersion, and $\Sigma$ is the mass surface density. Before combining the three values, we continue to follow \citet{2013MNRAS.433.1389R} and calculate corrective weighting terms to incorporate the effects of disk thickness,
\begin{equation}
T_i=\begin{cases}
1+0.6\left(\displaystyle\frac{\sigma_{z,i}}{\sigma_{R,i}}\right)^2\ \textrm{for}\ 0\leq\sigma_{z,i}/\sigma_{R,i}\leq0.5,\\
0.8+0.7\left(\displaystyle\frac{\sigma_{z,i}}{\sigma_{R,i}}\right)\ \textrm{for}\ 0.5\leq\sigma_{z,i}/\sigma_{R,i}\leq1,\\
\end{cases}
\end{equation}where $\sigma_{z,i}$ and $\sigma_{R,i}$ are the vertical and radial velocity dispersions for each species $i$. Following \citet{2013MNRAS.433.1389R}, we also compute the weight factor
\begin{equation}
w_i = \frac{2 \sigma_m \sigma_i}{\sigma^2_m + \sigma^2_i},
\end{equation} for each species $i$, where $m$ is the species with the smallest $Q_iT_i$. The combined value of $Q_\mathrm{RF}^{\phantom i}$ is calculated through both of these weighting factors,
\begin{equation}
\frac{1}{Q_\mathrm{RF}^{\phantom i}} = \sum_i \frac{w_i}{T_iQ_i},
\end{equation} where the sum is over all species $i$. We then calculate the characteristic instability wavelength as
\begin{equation}
\lambda_\mathrm{RF}^{\phantom i} = \frac{2\pi\sigma_m}{\kappa}.
\end{equation}

Note that $Q_\mathrm{RF}^{\phantom i}>1$ ensures stability against axisymmetric perturbations, while larger values of $Q$ ($\gtrsim2$) are required to stabilize the disk against non-axisymmetric perturbations \citep[e.g.][]{2012MNRAS.422..600G}. Note also that $\lambda_\mathrm{RF}^{\phantom i}$ is the scale at which the disk becomes unstable as $Q_\mathrm{RF}^{\phantom i}$ drops below unity. The usefulness of $Q_\mathrm{RF}^{\phantom i}$ and $\lambda_\mathrm{RF}^{\phantom i}$ as disk instability diagnostics has been illustrated by \citet{2015MNRAS.451.3107R,2016arXiv160203049R} and by \citet{2015ApJ...806L..34F}.

We calculate the values in of $Q_\mathrm{RF}^{\phantom i}$ and $\lambda_\mathrm{RF}^{\phantom i}$ in $80$ pc radial bins at each output dump, and then calculate the area-weighted geometric means across two zones -- the inner disk zone where most of the star formation happens, and the intermediate disk zone. These are plotted in Figure~\ref{qtime}.

As the simulation evolves and gas becomes increasingly centrally concentrated and then consumed by star formation, $Q$ and $\lambda$ become large, but there is a trend in the outer disk for $\lambda$ and $Q$ to increase as the strength of tides increases. In E, tidal stripping results in a dearth of particles in the intermediate disk zone, and $Q$ and $\lambda$ become noisy, but still generally keep to this trend. More critically, the inner disk $Q$ before $t=1$ Gyr appears to increase with the star formation rate -- B and C have a lower $Q$ than A, which in turn has a lower $Q$ than D and E. This trend is more visible in Table~\ref{qsfr}, where we calculate $\langle Q \rangle$, the geometric mean across the inner disk $Q$ calculated across all dumps in first $0.75$ Gyr, and find that $\langle Q \rangle$ increases with the total mass of star formation.

There are two potential interpretations for this. The correlation of a large $Q$ with strong star formation seems to suggest that bar and Toomre instabilities are unlikely to be the cause of this inflow of gas, as these effects should be strongest when $Q$ is low. In this case, it is possible that the inflow and the star formation it induces is produced by the effective viscosity of a disk with a large velocity dispersion. Figure~\ref{rothistvr} shows that feedback and tides can produce inflows and outflows with velocities of $|v|>10$ km~s$^{-1}$. The gas radial velocity dispersion (Figure~\ref{rothistdv}) can also reach above $10$ km/s. With circular velocities of $\sim20-40$ km/s (Figure~\ref{rothistvc}), this means the gas disks are extremely violent. While we do not resolve a full turbulent cascade, such a tumultuous disk can transfer mass and angular momentum and dissipate kinetic energy in a manner analogous to a turbulent viscosity, concentrating gas in the centre of the galaxy \citep{1981ARA&A..19..137P}. This could also explain why the inner disk $Q$ increases with the star formation rate. When the velocity dispersion is larger, the disk is more stable against Toomre instabilities and $Q$ is larger as well. The effective viscosity increases in efficiency with the velocity dispersion, with the result that, for these violent disks, gas becomes centrally concentrated more rapidly and produces a larger burst of star formation when $Q$ is large. However, this does not explain why the amplitudes of the Fourier modes also correlate with the star formation rate.
Alternatively, the large values of $Q$ could be a {\em result} of significant instability, which is likely to be the dominant source of turbulence and transport in galactic disks, as argued by \citet{2015ApJ...814..131G,2016arXiv160500646G}. Gravitational instabilities can stir the disk, and so the larger $Q$ is a result of a velocity dispersion produced by stronger gravitational instabilities. This would explain why the amplitudes of the Fourier modes, the value of $Q$, the mass of star formation, and the strength of tidal forces all increase together. This is the more likely conclusion, as it can explain all of these trends simultaneously. Hence, gravitational instabilities appear to be a significant factor in determining the star formation rate, and strong tides appear to play a significant role in the strength of these instabilities.

\makeatletter{}\subsection{Tidal stripping}\label{stripsection}

\begin{figure*}
\centering
\includegraphics[width=.98\textwidth]{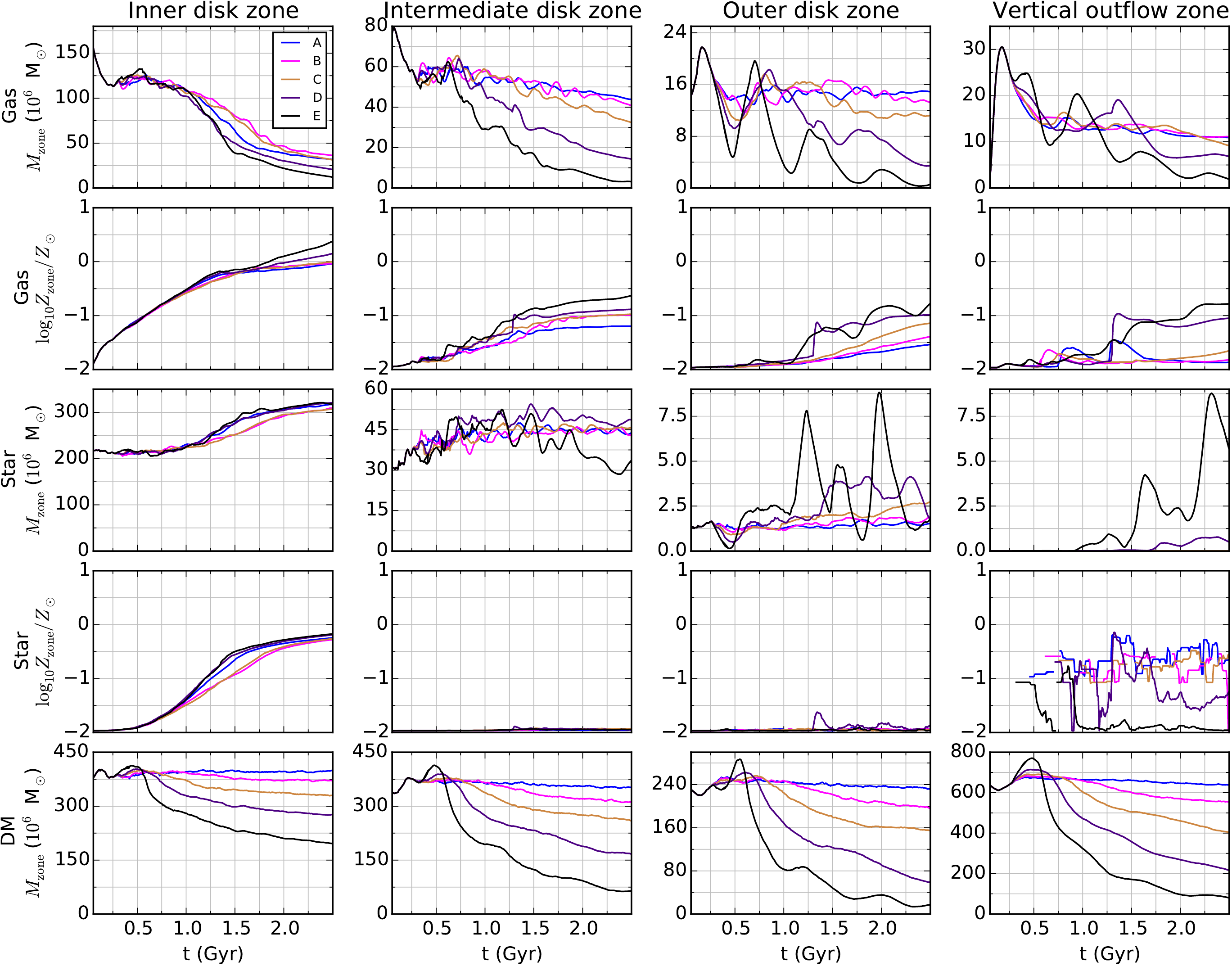}
\caption{\label{zones}
Evolution of the mass of stars, gas, and dark matter, and metallicity of stars and gas, in different zones of the disk. Here the stellar mass and metallicity describe the entire population of stars (both formed and initial stars).
}
\end{figure*}

To follow the evolution of the mass distribution (and hence, tidal stripping), we have divided the domain of the simulation into zones, and track the evolution of mass and metallicity in the zones defined in section~\ref{nomensec}. The mass of gas, stars, and dark matter, and the metallicity of gas and stars in each zone as a function of time are plotted in Figure~\ref{zones}.

In the inner disk zone, star formation is the dominant process, reducing the gas mass, increasing the stellar mass, and increasing the mean metallicity of gas and stars. There is a small increase in the inner disk zone gas mass at $t\approx0.5$ Gyr resulting from the inflow of gas, but star formation dominates in the long term. The differences in the gas or stellar mass between the simulations is largely consistent with the differences in the total mass of star formation, as shown in Table~\ref{qsfr} -- tidal stripping is weak in the inner disk zone.
In the intermediate disk zones, the gas mass decreases over time, with a rate that becomes more rapid with increasing tide strength. It is not clear from Figure~\ref{zones} alone whether this gas has been depleted by radial inflow, tidal stripping, or star formation, but by tracking the individual particles that are in the intermediate disk zone at $t=0.5$ Gyr, we confirmed that the primary source of mass loss in this zone is from gas particles leaving the disk. However, at the same time there is comparatively little change in the stellar mass of the intermediate disk zone. This suggests that the gas mass loss is not caused by the purely dynamical effects of the external tidal fields, but is a combination of hydrodynamic and gravitational effects. That is, the tidal field enhances the ability of feedback to remove gas from the galaxy, but this field is not strong enough in itself to remove stars from this region.

The outer disk zone shows a greater sensitivity to tidal forces. Here we see a periodic effect when tides are strong, where mass flows in and out of the zone as the dwarf galaxy orbits through the halo and changes its relative orientation. This effect is present even in the stellar mass, but the stars in this region make up only $<0.1\%$ of the stellar disk, and hence the bulk of the stars are not significantly stripped by tidal forces.

We can further quantify the effects of tidal stripping by measuring the radial scale length of the gas disks. As noted in section~\ref{nomensec}, our method for measuring the scale length is evenly weighted over $20$ kpc, and hence is dominated by the kpc-scale structure of the outer disk. This makes it a good measure of tidal stripping of the outer disk that is not strongly affected by the concentration of gas in the inner disk. The results are given in Table~\ref{qsfr}. Here we see a clear trend of the gas disk scale length decreasing with increasing tide strength. The tides contribute to removing the outer material, producing a truncated gas disk.

We find that dark matter (bottom row of Figure~\ref{zones}) is stripped in all zones in all runs B-E, even within the central $2$ kpc, despite the tidal radius being much larger than this radius, ranging from $\sim7-22$ kpc. This is caused by the eccentricity of the orbits of the dark matter particles. Particles with a periapsis near the centre of the dwarf galaxy will often have an apoapsis that is beyond the tidal radius, allowing them to be stripped. Nevertheless, stripping clearly becomes stronger as we move to outer parts of the dwarf.

\makeatletter{}\subsection{Metallicity Gradients}\label{flatsection}

\begin{figure*}
\begin{center}
\includegraphics[width=.98\columnwidth]{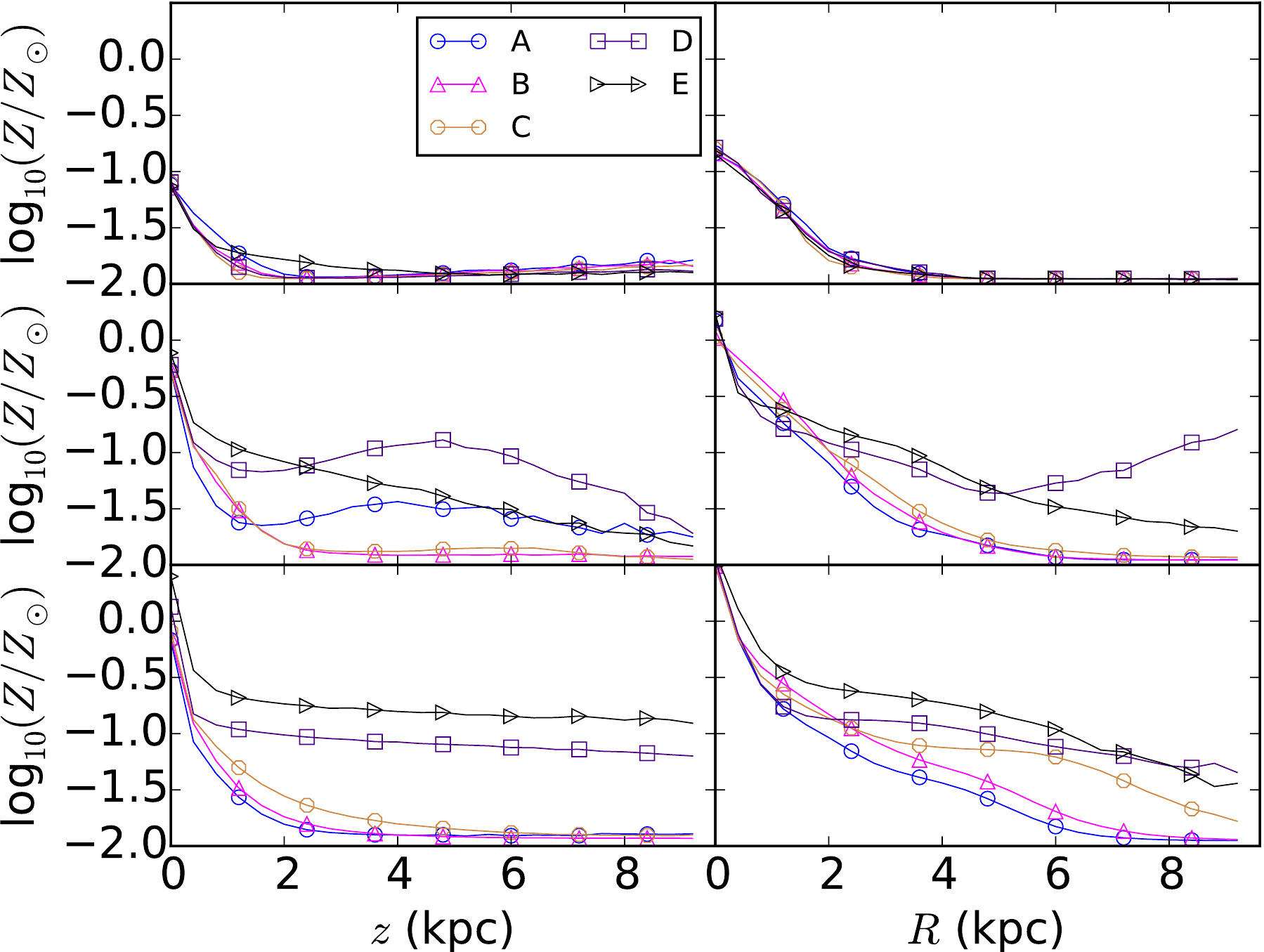}
\includegraphics[width=.98\columnwidth]{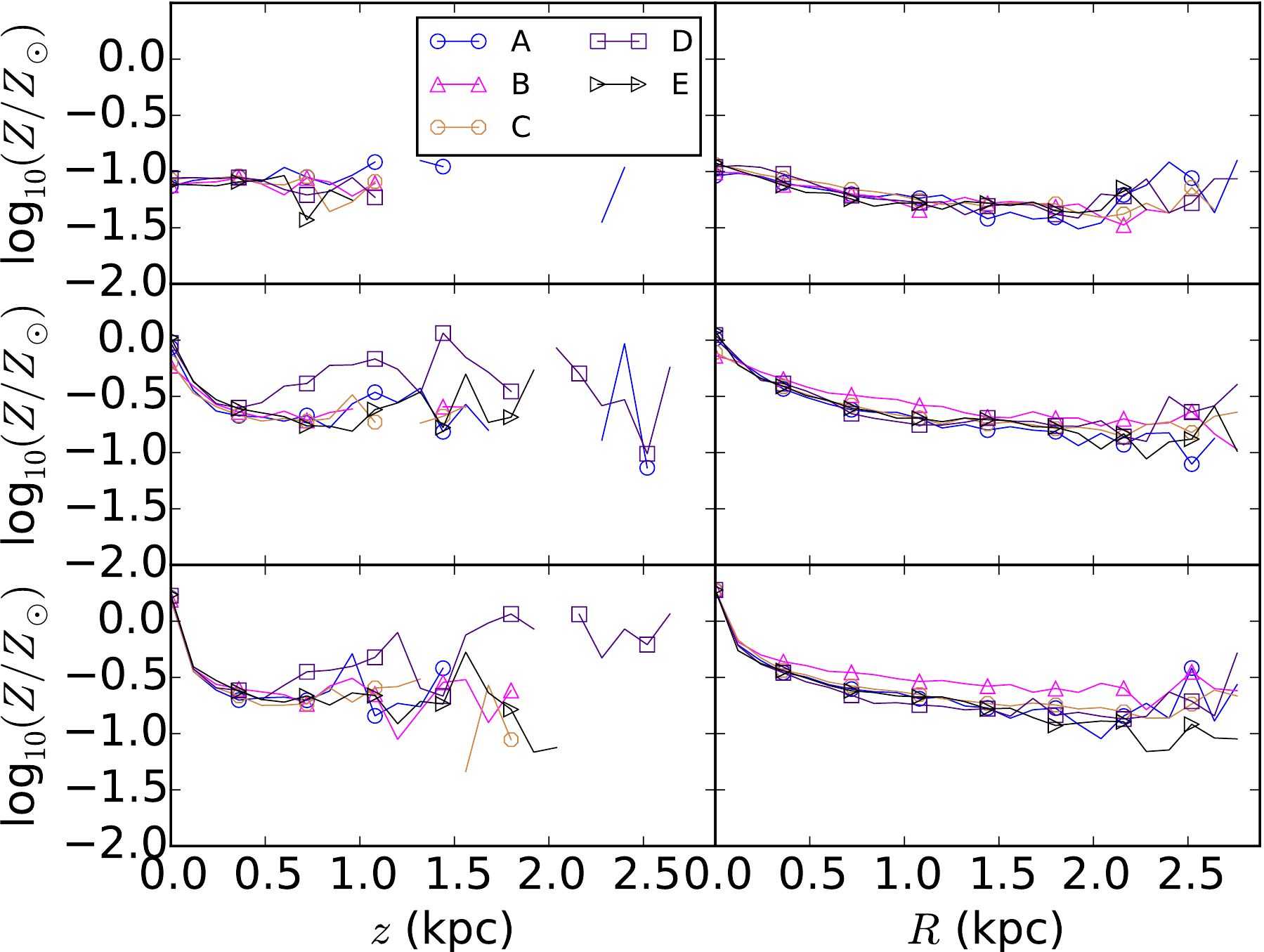}\\
\end{center}
\caption{\label{metslopes}
Metallicity profiles at $t=0.5$ Gyr (top), $t=1.5$ Gyr (centre), and $t=2.5$ Gyr (bottom) for gas (left two columns) and stars (right two columns). The stellar metallicity plots only include stars formed during the simulation. Note that the stellar and gas plots span different ranges of $R$ and $z$.
}
\end{figure*}

\begin{figure}
\begin{center}
\includegraphics[width=.98\columnwidth]{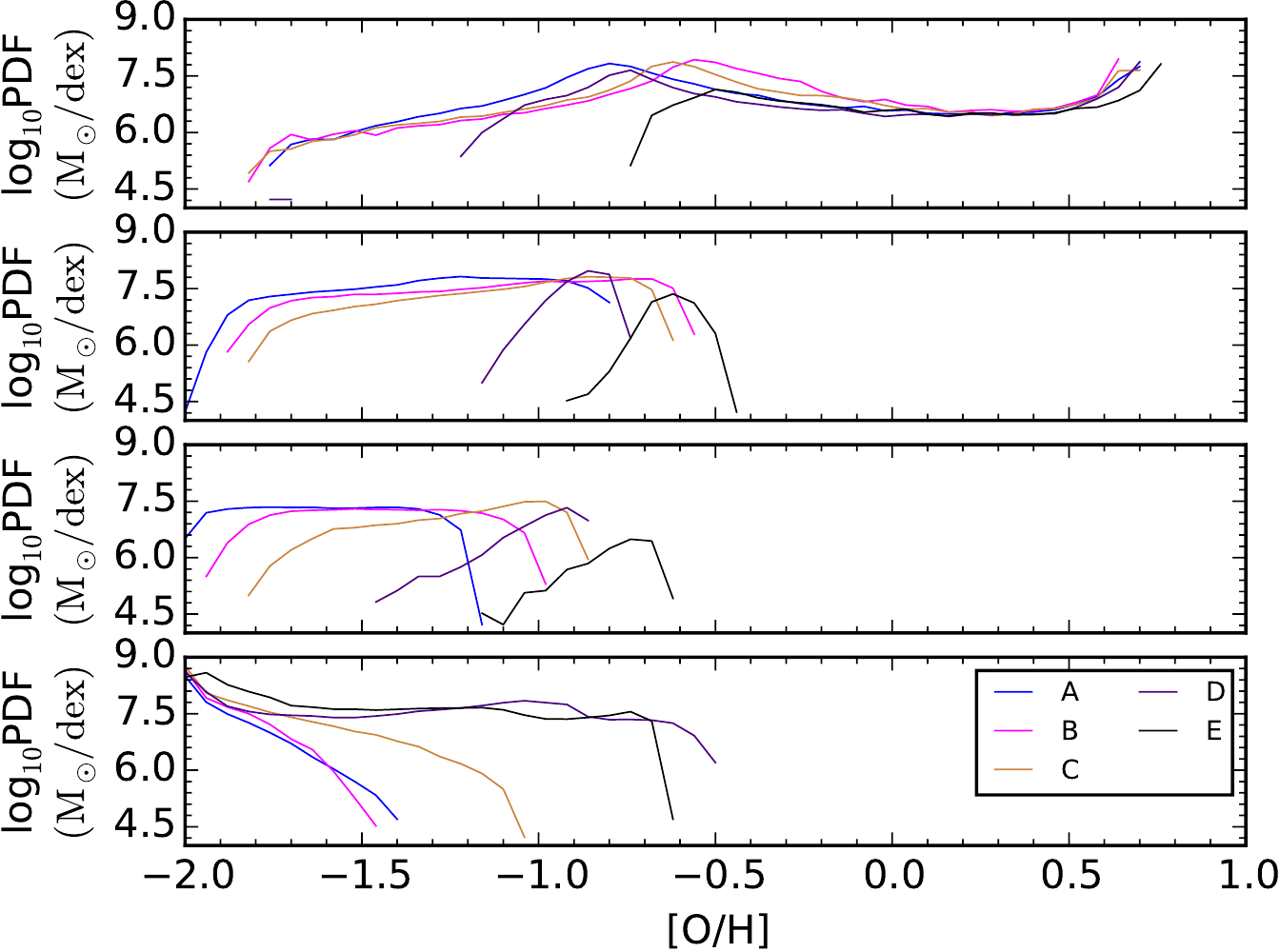}
\end{center}
\caption{\label{gasmdfs}
Gas MDFs at $t=2.5$ Gyr in the different disk zones. From top to bottom: inner disk zone; intermediate disk zone; outer disk zone; vertical outflow zone.
}
\end{figure}

The metallicity profiles at $t=0.5$ Gyr, $t=1.5$ Gyr, and $t=2.5$ Gyr for the gas and stellar components are plotted in Figure~\ref{metslopes}. The stellar metallicity profile only incorporates stars that were formed during the simulation, as the stars present in the initial conditions all have a constant metallicity.

In all simulations, the centrally concentrated star formation produces a significant negative metallicity slope, both vertically and horizontally, in both the gas and stellar components. Over time, the central metallicity continues to increase as star formation proceeds, and this metallicity can spread outwards through diffusion, feedback-driven bulk outflows, and tidal effects.

Hydrodynamic simulations have tended to show flatter metallicity gradients for low-mass galaxies \citep[e.g.][]{2003MNRAS.340..908K,2011MNRAS.411..627T}, but observations show that significant metallicity gradients can exist in dwarf galaxies \citep{2009AJ....137.3100W,2009AN....330..960K,2011MNRAS.417.1643K}, and that the existence and even sign of any correlation between metallicity gradients and galaxy masses is still disputed \citep[e.g.][]{2005MNRAS.361L...6F,2007A&A...463..455A,2009ApJ...705..723C,2011MNRAS.417.1643K,2014MNRAS.442.1003P,2015ApJ...808...26R}.

As noted in section~\ref{stripsection}, tidal effects are weak by themselves, and differing strengths of tides do not appear to have a significant effect on the stellar metallicity profile. Stars are formed within the central region ($R\lesssim2$ kpc, $z\lesssim2$ kpc), and do not migrate very efficiently, producing a sharp central peak in the metallicity slope that reflects the concentrated metallicities of the gas distributions.

The gas metallicity profiles show much clearer effects. At $t=2.5$ Gyr, there is a trend in both the horizontal and vertical profiles for models with stronger tides to have flatter slopes, with a higher metallicity at larger distances from the centre. However, feedback-driven outflows can cause significant short-term deviations from this trend. At $t=1.5$ Gyr, A and D show positive vertical metallicity gradients for at least part of the gas disk, and demonstrate that metals are at least temporarily more spatially distributed than in simulations with stronger tides. It is not likely a coincidence that this occurs near when the star formation rates reach their peaks (Figure~\ref{SFR}). Nevertheless, these are short-term variations, and the general trend is to have flatter gradients with stronger tides.

It is important to determine whether these flatter metallicity gradients are simply a result of the disk spreading out due to the tidal potential and the stripping of the dark matter halo. In other words, we should determine whether the metallicity slope in units of dex per scale length would show as strong a trend as the slope in units of dex per kpc. This can be done by examining the scale lengths of the gas disks, as calculated in section~\ref{stripsection} and plotted in Table~\ref{qsfr}. The scale lengths {\em decrease} with increasing tide strength (that is, with steeper metallicity gradients), indicating that the trend of steeper metallicity gradients with stronger tides would be {\em amplified} if taken in units of dex per scale length. Hence, this trend is robust against this change of definition.

The metallicity density functions (MDFs) for gas at $t=2.5$ Gyr across the different disk zones are plotted in Figures~\ref{gasmdfs}, using [O/H] as a measure of the metallicity. 
The MDFs show the effects of tidal stripping, with the lower cutoff of the inner, intermediate, and outer disk MDFs moving to greater metallicities as the strength of tides increases. Conversely, the upper cutoff of the vertical outflow zone, which shows gas that has escaped from the disk proper, moves to higher metallicities as the strength of tides increases. The highest metallicity gas remains only in the inner disk zone, even when tides are strong. This confirms our earlier conclusion that gas does not efficiently spread out from the inner regions of the disk to the outer regions, but is instead completely stripped from the galaxy. From the disk and vertical outflow zones, we see that it is mostly the low-metallicity gas that is stripped, producing a higher-metallicity remnant disk. As the strength of tides increases, gas with increasingly higher metallicity is stripped, producing a remnant with an increasingly higher mean metallicity.

\makeatletter{}\subsection{Comparison with mass-metallicity relation}\label{massmetsection}

\begin{table}
\begin{center}
\begin{tabular}{cccccc}
\hline\hline
~ & ~ & $12+\log(\mathrm{O}/\mathrm{H})$ & ~ & ~ & ~\\
Run & (inner disk) & (disk) & (all) & $\log y_\mathrm{eff}$ & $\log M_*$\\
\hline
A &$8.95 $&$8.50$&$ 8.35 $&$ -2.38$&$8.56$\\
B &$8.96 $&$8.60$&$ 8.42 $&$ -2.31$&$8.55$\\
C &$8.99 $&$8.64$&$ 8.39 $&$ -2.30$&$8.55$\\
D &$9.12 $&$8.89$&$ 8.42 $&$ -2.19$&$8.57$\\
E &$9.39 $&$9.27$&$ 8.40 $&$ -1.95$&$8.55$\\
\hline
\end{tabular}
\end{center}
\caption{\label{massmetab} \textup{Gas-phase metallicity of the inner disk zone, for the entire disk ($|z|\leq2$ kpc, $R\leq10$ kpc), and for all gas particles, the effective yield ($y_\mathrm{eff}$), and total stellar mass ($M_*$) for all runs at $t=2.5$ Gyr
}
}
\end{table}

The metallicity observations of \citet{2004ApJ...613..898T} are typically measurements of the central metallicity, as their median projected fiber size has a diameter of $4.6$ kpc. Our inner disk zone has a comparable diameter of $4$ kpc, and so for comparison we have calculated the metallicity of gas in the inner disk zone of each of our galaxies at $t=2.5$ Gyr. To measure the relative importance of tidal stripping and star formation, we also calculate the gas-phase metallicity of all disk gas ($|z|\leq2$ kpc, $R\leq10$ kpc), and the mean metallicity of {\em all} gas in the simulation, including gas that has completely escaped the dwarf galaxy. We have tabulated these metallicities, along with the total stellar disk mass of both formed and initial stars, and with the effective yield (detailed below), in Table~\ref{massmetab}. Here we use $12+\log(\mathrm{O}/\mathrm{H})$ as a measure of the metallicity, to facilitate comparison with \citet{2004ApJ...613..898T}.

The total stellar disk mass shows no correlation with tide strength, but there is a clear trend of stronger tides producing greater metallicities in the inner disk zone, and across the disk in general. The disk and inner disk metallicities increase monotonically with tide strength, unlike in Table~\ref{qsfr}, where the metal production slightly decreases with weak tides, and unlike the mean metallicity of all gas. The mean metallicities of all gas differ by 0.07 dex at most, while the inner disk metallicities differ by 0.46 dex. This demonstrates that the differences in star formation rates resulting from the influence of tides on gravitational instabilities are not the key factor in driving the central metallicity. Instead, tidal stripping plays the primary role, by stripping low-metallicity gas from the galaxy. The metallicity gradient also contributes, giving an inner disk metallicity that is greater than the entire disk metallicity.

To further quantify the effects of tidal stripping, we calculate the effective yield as in \citet{2004ApJ...613..898T} by inverting the equation for metallicity evolution in the closed-box instantaneous single-zone approximation,
\begin{equation}
Z=y\ln(\mu^{-1}),
\end{equation}where $\mu$ is the gas mass fraction, $y$ is the stellar yield, and $Z$ is the gas-phase metallicity, to calculate the effective yield
\begin{equation}
y_\mathrm{eff}=Z/\ln(\mu^{-1}).
\end{equation}We tabulate these values at $t=2.5$ Gyr in Table~\ref{massmetab}, using the gas fraction and metallicity of the entire disk for $Z$ and $\mu$. We use the total metallicity of all species here instead of just the oxygen metallicity to capture the effects of all metals that escape, but we note that oxygen mass fraction of disk gas metals is nearly constant between simulations at this time, ranging from $0.56$ to $0.61$. The clear result is that the effective yield increases with increasing tide strength, confirming that it is tidal stripping, and not star formation, that drives the enhanced metallicities.

It may seem unusual that the all-gas metallicities of runs B \& C are slightly higher than those of run A, despite run A producing more metals than runs B \& C (Table~\ref{qsfr}). This is possible because our simulations are not bound by the closed-box instantaneous single-zone approximation. Hence, star formation can -- at least temporarily -- reduce the gas-phase metallicity if stars form from the highest metallicity gas, especially for recently formed star particles that have not had sufficient time to return their metal yield to the ISM, or if large quantities of gas have escaped the disk and are unable to be enriched.

The inner disk metallicities of our simulated galaxies are somewhat large, with even model A at the upper range for galaxies of this mass \citep{2004ApJ...613..898T}. This may be a result of the feedback model used here. Simulations performed at higher resolution \citep{2014MNRAS.438.1208K}, or with localized FUV radiation from star formation \citep{2012MNRAS.421.3488H,2016arXiv160500650F} could perhaps maintain a warmer or more disturbed disk with a lower star formation rate. However, some dwarf galaxies (i.e. BCDs) are also known to have low metallicities \citep{1972ApJ...173...25S,1994ESOC...49..421T,2012AJ....144..134H} despite significant star formation rates \citep{1988ApJ...334..665F}, which may imply that allowing metals to escape the dwarf through galactic winds may be more critical than reducing the star formation rate. The finite mass resolution of SPH simulations makes winds difficult to resolve, but could perhaps be captured by higher resolution simulations with a long-range radiation pressure model \citep{2012MNRAS.421.3522H}.

Nevertheless, from our results we have found evidence that tides cause galaxies to move to higher metallicities in the mass-metallicity diagram. This may be one of the causes of the intrinsic scatter around the mass-metallicity relation.

\makeatletter{}\section{Discussion}\label{discsection}

Our primary result is that the metallicity of dwarf galaxies can be enhanced by the presence of a tidal field, due to centralized star formation producing a metallicity gradient, and tidal stripping removing the outer low-metallicity gas. The strength of this centralized star formation is also affected by gravitational instabilities, which can be strengthened by strong tides.

We have produced these results with hydrodynamic simulations of comparatively massive dwarf disk galaxies, using a single set of reasonable initial conditions and physical parameters. Here, we discuss the robustness of our results against changes to these values. Given the large number of degrees of freedom, it would require a very large number of simulations to completely map the parameter space. However, there exists a number of previous studies that explore variations to the relevant parameters, and we use these to produce a series of predictions.

Firstly, we examine factors that could influence the strength of tidal stripping. We should expect the eccentricity and semi-major axis of the orbit to have a significant effect. Tides become much more effective as a dwarf approaches the core of the host halo, and orbits that plunge deep into the halo experience particularly strong tidal forces \citep[e.g.][]{2001ApJ...559..754M,2010MNRAS.405.1723S,2011MNRAS.415.1783B,2011ApJ...726...98K,2015MNRAS.454.2502S}.  The orientation of the dwarf galaxy against its host halo may affect the effectiveness of tidal stripping and stirring, being suppressed under retrograde rotation \citep[e.g.][]{2001ApJ...547L.123M,2001ApJ...559..754M,2006MNRAS.366..429R,2012MNRAS.424.2401V}. Hence, the strength of stripping will depend on the parameters of the orbit.

The mass and scale-length of the dwarf disk should also affect the effectiveness of tides. Tidal stripping becomes more effective as the mass of the dwarf galaxy is reduced \citep{2011ApJ...726...98K} for a fixed orbit. On the other hand, dynamical friction, which we have not included, becomes more significant as mass increases, causing more massive galaxies to more rapidly reach the centre of the halo where tidal effects are strong \citep{2012MNRAS.424.2401V}. Tidal stripping also becomes more effective as the scale-length of the dwarf is increased \citep{2011ApJ...726...98K}.

Secondly, we examine factors that could influence the production of a metallicity gradient. Tidal stripping removes the outer material from the dwarf galaxy. Hence, when the metallicity gradient is steep, tidal stripping primarily removes material with very low metallicities, producing a truncated metal-rich disk. When the metallicity gradient is shallow, tidal stripping removes a larger fraction of the disk's metallicity, and the resulting truncated disk is less enriched.

The thickness of the disk should not greatly affect the strength of tidal stripping \citep{2012MNRAS.424.2401V}, but could somewhat affect the strength of gravitational instabilities \citep{1992MNRAS.256..307R,2011ApJ...737...10E}, and hence the star formation rate. This could result in a steeper or shallower metallicity gradient.

The gas fraction of the disk, and its mass and scale length will also affect star formation and the formation of gravitational instabilities. The Kennicutt-Schmidt \citep{1959ApJ...129..243S,1998ApJ...498..541K} law demonstrates that the star formation surface density increases with the gas surface density. This means that the total star formation rate will increase with the mass and gas fraction of the disk, but decrease with the scale length. Slower star formation will result in a gentler metallicity gradient. Gravitational instabilities will also become stronger as the mass and surface density of the dissipational component is increased \citep{2011ApJ...737...10E}, and so we might expect stronger instabilities that drive gas inwards more rapidly with larger mass, larger gas fraction, and shorter scale-lengths. However, we should still expect star formation to be concentrated in the centre, as is observed in many dwarf galaxies \citep{2008ApJ...682L..89G,2009ApJ...703..692L,2011MNRAS.417.1643K,2012A&A...539A.103D,2013ApJ...778..103H}, and hence a metallicity gradient will still be produced, provided that radial mixing is inefficient.

As noted above, the strength of diffusion in SPH simulations of galaxies is difficult to constrain. Diffusion in this regime cannot be easily measured by experiment, and cannot be calibrated by direct comparison with observations, as the distribution of metals depends on many other factors that are often model-dependent, such as the star formation and feedback algorithm. We investigated the effects of different strengths of diffusion in dwarf galaxy simulations in Paper \textsc{I}. Here, we found that the distribution of metals only has a weak dependence on the strength of diffusion, provided that some diffusion is present. In the complete absence of diffusion, there is an over-abundance of low-metallicity stars, but even weak diffusion removes this. With stronger diffusion, the peak of the stellar metallicity MDF is pushed to lower metallicities. The metallicity gradient also becomes steeper with stronger diffusion, as metals are diffused out of outflows, reducing the effectiveness of metal mixing through the bulk flow of gas. However, these are weak effects. The proportion of metals entrained in an outflow increased by a factor of $50$ \% when the diffusion coefficients were reduced by a factor of $10$. Hence, even large variations in the strength of diffusion should not greatly change our results.

The implementation and choice of parameters for feedback may also affect the metallicity gradient. Stronger feedback can produce a flatter gradient \citep{2013A&A...554A..47G} by more efficiently distributing metals over large scales.

Altogether, variations in these parameters should change the strength of the important processes - gravitational instability, centralized star formation, and the stripping of outer material. However, for modest variations we do not expect the basic qualitative results to change: that centralized star formation (perhaps enhanced by gravitational instability) and the tidal stripping of outer material combine to produce a galaxy that is truncated and richer in metals than an unstripped disk.

\section{Conclusions}\label{concsection}
We have performed simulations of dwarf galaxies under the influence of tidal fields to determine the effects on the chemodynamical evolution of dwarfs and the implications to the mass-metallicity relation. We have found that

\begin{enumerate}
\item all of the dwarf galaxies undergo gravitational instabilities that drive gas inwards, resulting in rapid central star formation and a metallicity gradient;
\item strong tides strengthen these instabilities, increasing the star formation rate and slightly increasing the metallicity; 
\item with this metallicity slope, tidal stripping preferentially removes the outer low-metallicity gas, producing a truncated gas disk with significantly higher metallicity and a shallower metallicity gradient
\end{enumerate}

Most significantly, in these simulations we find that tides do not contribute to the lack of metals in low-mass galaxies, but instead {\em increase} their metallicities, primarily through stripping of low-metallicity gas. Hence, as tides cause galaxies to move to {\em higher} metallicities in the
mass-metallicity relation, tides do not help to explain the mass-metallicity relation except as a contribution to its scatter. However, we note that these results may be sensitive to resolution, orbital parameters, initial conditions, and the treatment of feedback.

\section*{Acknowledgements}
We thank our referee for their detailed comments that have helped to strengthen this paper. This research was supported by the Canada Research Chair program and NSERC. Simulations were run on the Calcul Qu\'{e}bec/Compute Canada supercomputers {\em Colosse} and {\em Guillimin}. We thank Daisuke Kawata for use of the \textsc{GCD+} code and for useful comments that contributed to this paper.\\

\bibliographystyle{apj}
\bibliography{fulltext_tides1}

\end{document}